\documentclass[12pt]{article}
\usepackage{epsfig,psfrag,cite,float}
\def \ra{\rightarrow}

\def \beq{\begin{equation}}
\def \eeq{\end{equation}}
\def \bea{\begin{eqnarray}}
\def \eea{\end{eqnarray}}
\def \ben{\begin{enumerate}}
\def \een{\end{enumerate}}
\def \bit{\begin{itemize}}
\def \eit{\end{itemize}}

\def \branch{{\cal B}}
\def \Im{{\hbox{Im}}\,}
\def \Re{{\hbox{Re}}\,}
\def \gev{{\hbox{GeV}}}

\def \mev{{\hbox{MeV}}}
\def \tev{{\hbox{TeV}}}

\def \cl#1{{#1\%\ \mathrm{C.L.}}}

\def \eq#1{Eq.~(\ref{#1})}

\def \fig#1{Fig.~\ref{#1}}

\def \nn{\nonumber}

\def \a{\alpha}
\def \b{\beta}
\def \D{\Delta}
\def \g{\gamma}

\def \d{\delta}
\def \e{\epsilon}

\def \l{\lambda}

\def \m{\mu}
\def \n{\nu}

\def \p{\pi}
\def \r{\rho}
\def \s{\sigma}

\begin{document}
\begin{titlepage}

\begin{flushright}
DESY 01-059\\
hep-ph/0105200\\
May 2001\\
\end{flushright}
\begin{center}
\LARGE
Extended Minimal Flavour Violating MSSM and \\
Implications for $B$ Physics
\end{center}

\bigskip

\begin{center}
\Large
A. Ali\footnote{E-mail address: ali@x4u2.desy.de} ~and     
E. Lunghi\footnote{E-mail address: lunghi@mail.desy.de}
\end{center}

\begin{center}
\large
Deutsches Elektronen Synchrotron, DESY, \\
Notkestrasse 85, D-22607 Hamburg, Germany
\end{center}

\bigskip
\begin{abstract}
The recently reported measurements of the CP asymmetry $a_{\psi K}$ by
the BABAR and BELLE collaborations, obtained from the rate differences
in the decays $B^0 \to (J/\psi K_s), (J/\psi K_L)$ etc., and their
charge conjugates, are in good agreement with the standard model (SM)
prediction of the same, resulting from the unitarity of the CKM
matrix. The so-called minimal flavour violating (MFV) supersymmetric
extensions of the standard model, in which the CKM matrix remains the
only flavour changing structure, predict $a_{\psi K}$ similar to the
one in the SM. With the anticipated precision in $a_{\psi K}$ and
other CP asymmetries at the B factories and hadron colliders, one
hopes to pin down any possible deviation from the SM. We discuss an
extension of the MFV-supersymmetric models which comfortably
accommodates the current measurements of the CP asymmetry $a_{\psi
K}$, but differs from the SM and the MFV-supersymmetric models due to
an additional flavour changing structure beyond the CKM matrix. We
suggest specific tests in forthcoming experiments in $B$ physics. In
addition to the CP-asymmetries in $B$-meson decays, such as $a_{\psi
K}$ and $a_{\pi \pi}$, and the mass difference $\Delta M_s$ in the
$B_s^0$ - $\overline{B_s^0}$ system, we emphasize measurements of the
radiative transition $b \to d \gamma$ as sensitive probes of the
postulated flavour changing structure. This is quantified in terms of
the ratio $R(\rho \gamma/K^* \gamma) = 2{\cal B}(B^0 \to \rho^0
\gamma)/{\cal B}(B^0 \to K^{* 0} \gamma)$, the isospin violating ratio
$\Delta^{\pm 0}={\cal B}(B^\pm \to \rho^\pm \gamma)/2{\cal B}(B^0 \to
\rho^0 \gamma) -1$, and the CP-asymmetry in the decay rates for $B^+
\to \rho^+ \gamma$ and its charge conjugate. Interestingly, the
CKM--unitarity analysis in the Extended--MFV model also allows
solutions $\bar\rho <0$ for the Wolfenstein parameter, as opposed to
the SM and the MFV-supersymmetric models for which only $\bar\rho > 0$
solutions are now admissible, implying $\gamma > \pi/2$, where
$\gamma=-\arg V_{ub}$. Such large values of $\gamma$ are hinted by the
current measurements of the branching ratios for the decays $B\to
\pi\pi$ and $B\to K \pi$.
\end{abstract}
\end{titlepage}

\section{Introduction}
With the advent of the B-factory era, the principal focus in flavour
physics is now on measuring CP-violating asymmetries, which will
determine the inner angles $\alpha$, $\beta$, and $\gamma$ of the
unitarity triangle (UT) in the Cabibbo-Kobayashi-Maskawa (CKM)
theory~\cite{ckm}. A beginning along this road has already been made
through the impressive measurements of $\sin 2 \beta$ by the B-factory
experiments BABAR~\cite{Aubert} and BELLE~\cite{Abashian}, following
earlier leads from the OPAL~\cite{Ackerstaff},
CDF~\cite{Affolder,Blocker}, and ALEPH~\cite{Barate}
collaborations. The principal decay modes used in the measurement of
$\sin 2 \beta$ are $B_d^0 \to J/\psi K_S$, $B_d^0 \to \psi (2S) K_S$,
$B_d^0 \to J/\psi K_L$, and their charge conjugates. Concentrating on
the decays $B_d^0/\overline{B_d^0} \to J/\psi K_s$, the time-dependent
CP-asymmetry $a_{\psi K_S}(t)$ can be expressed as follows: \bea
a_{\psi K_S}(t) &\equiv& \frac{\Gamma(B_d^0 (t) \to J/\psi K_S) -
\Gamma(\overline{B_d^0}(t) \to J/\psi K_S)} {\Gamma(B_d^0 (t) \to
J/\psi K_S) + \Gamma(\overline{B_d^0}(t) \to J/\psi K_S)} \nn\\
&=&{\cal A}_{\rm CP}^{\rm dir} \cos(\D M_{B_d} t) + {\cal A}_{\rm
CP}^{\rm mix} \sin (\D M_{B_d} t) \; ,
\label{apsiksdef}
\eea
where the states $B_d^0(t)$ and $\overline{B_d^0}(t)$ are understood as
evolving from the corresponding initial flavour eigenstates (i.e., at $t=0$),
and $\D M_{B_d}$ is the mass difference between the two mass eigenstates
of the $B_d^0$ -$\overline{B_d^0}$ system, known very
precisely, thanks in part due to the BABAR \cite{Bozzi2001} and BELLE
\cite{Abe2000} measurements, and the present world average is $\D M_{B_d}=
0.484 \pm 0.010$ (ps)$^{-1}$
\cite{HFWG}. The
quantities ${\cal A}_{\rm CP}^{\rm dir}$ and ${\cal A}_{\rm CP}^{\rm mix}$
are called the direct (i.e., emanating from the decays) and mixing-induced
CP-asymmetries, respectively. Of these, the former is CKM-suppressed - a 
result which holds in the SM. The expectation
${\cal A}_{\rm CP}^{\rm dir}/{\cal A}_{\rm CP}^{\rm mix} \ll 1$
is supported by present data on direct CP-asymmetry in 
charged B-decays, $B^\pm \to J/\psi K^\pm$, yielding an upper bound on
${\cal A}_{\rm CP}^{\rm dir}$ which is already quite stringent
\cite{Aubert,Abashian}. Hence, we shall assume that direct CP-asymmetry in
$a_{\psi K_S}(t)$ is negligible and neglect the first term on the
r.h.s. of Eq.~(\ref{apsiksdef}). Recalling that ${\cal A}_{\rm CP}^{\rm
mix}$ is a pure phase, one has in the SM  ${\cal A}_{\rm CP}^{\rm
mix}=\eta_{J/\psi K_S} a_{\psi K_S}$, with $\eta_{(J/\psi K_S)}=-1$
being the intrinsic CP-parity of the $J/\psi K_S$ state,
Eq.~(\ref{apsiksdef}) simplifies to 
\beq
a_{\psi K_S}(t) =- a_{\psi K_S} \sin (\D M_{B_d} t) ~.
\label{acpsinbeta}
\eeq 
This relation is essentially free of hadronic uncertainties. Hence, a
measurement of the left-hand-side allows to extract $a_{\psi K_S}$
cleanly. Note that in the SM $a_{\psi K_S} = \sin 2 \beta$.  The
present world average of this quantity is
\cite{Aubert,Abashian,Ackerstaff,Affolder,Blocker,Barate}
\beq 
a_{\psi K_S} =0.79 \pm 0.12 ~,
\label{sin2betawa}
\eeq 
which is dominated by the BABAR ($a_{\psi K_S} =0.59 \pm 0.14 {\rm
(stat)} \pm 0.05 ({\rm syst}))$ \cite{Aubert} and BELLE ($a_{\psi K_S}
=0.99 \pm 0.14 {\rm (stat)} \pm 0.06 ({\rm syst}))$ \cite{Abashian}
results.  We note that the current world average (which includes a
scale factor following the Particle Data Group prescription
\cite{Groom}) based on all five experiments yields a value of $a_{\psi
K_S}$ which is different from a null result by more than six standard
deviations.  To test the consistency of the SM, the current
experimental value of $a_{\psi K_S}$ in Eq.~(\ref{sin2betawa}) is to
be compared with the indirect theoretical estimates of the same
obtained from the unitarity of the CKM matrix. These latter values lie
typically in the range $a_{\psi K_S}^{\rm SM} =0.6 - 0.8$ (at 68\%
C.L.)  \cite{mele-ckm,ps-ckm,bargiotti-ckm,mfv,
ciuchini-ckm,buras-ckm,atwood,hocker-ckm}, where the spread reflects
both the uncertainties in the input parameters and treatment of
errors, with most analyses yielding $a_{\psi K_S}^{\rm SM} \simeq
0.70$ as the central value of the CKM fits. We conclude that he
current measurements of $a_{\psi K_S}$ are in good agreement with its
indirect estimates in the SM.

 The consistency of the SM with experiments on CP-violation in
B-decays will come under minute scrutiny, with greatly improved
accuracy on $a_{\psi K_S}$ and measurements of the other two angles of
the UT, $\alpha$ and $\gamma$ at the $e^+ e^-$ and hadronic
$B$-factories. In addition, a large number of direct CP-asymmetries in
charged and neutral $B$-decays, as well as
flavour-changing-neutral-current (FCNC) transitions in $B$- and
$K$-decays, which will be measured in the course of the next several
years, will greatly help in pinning down the underlying theory of
flavour physics.  It is conceivable that precision experiments in
flavour physics may force us to revise the SM framework by admitting
new interactions, including the possibility of having new CP-violating
phases.  Some alternatives yielding a lower value of $a_{\psi
K_S}$ than in the SM have already been entertained in the
literature~\cite{neubertbeta,wolfensteinbeta,nirbeta}. With the
experimental situation now crystallized in Eq.~(\ref{sin2betawa}), it now
appears that the CP-asymmetry  $a_{\psi K_S}$ has a dominantly SM origin. 

 In popular extensions of the SM, such as the minimal supersymmetric
standard model (MSSM), one anticipates supersymmetric contributions to
FCNC processes, in particular $\D M_{B_d}$, $\D M_{B_s}$ (the mass
difference in the $B_s^0$ -$\overline{B_s^0}$ system), and
$\epsilon_K$, characterizing ${\cal A}_{\rm CP}^{\rm mix}$ in the
$K^0$ -$\overline{K^0}$ system. However, if the CKM matrix remains
effectively the only flavour changing (FC) structure, which is the
case if the quark and squark mass matrices can be simultaneously
diagonalized (equivalently, the off-diagonal squark mass matrix
elements are small at low energy scale), and all other FC interactions
are associated with rather high scales, then all hadronic flavour
transitions can be interpreted in terms of the same unitarity
triangles which one encounters in the SM.  In particular, in these
theories $a_{\psi K_s}$ measures the same quantity $\sin 2 \beta$ as
in the SM. These models are usually called the minimal flavour
violating (MFV) models, following Ref.~ \cite{giudice}.  Despite the
intrinsic dependence of the mass differences $\D M_{B_d}$, $\D
M_{B_s}$, and $\epsilon_K$ on the underlying supersymmetric
parameters, the MFV models remain very predictive and hence they have
received a lot of theoretical attention lately
\cite{mfv,giudice,ali1,ali2,buras1,buras2,lunghi1,BF2001}.  To
summarize, in these models the SUSY contributions to $\D M_{B_d}$, $\D
M_{B_s}$, and $\epsilon_K$ have the same CKM-dependence as the SM top
quark contributions in the box diagrams (denoted below by
$C_1^{Wtt}$).  Moreover, supersymmetric effects are highly correlated
and their contributions in the quantities relevant for the UT-analysis
can be effectively incorporated in terms of a single common parameter
$f$ by the following replacement \cite{ali1,ali2}:
\beq
\e_K, \; \D M_{B_s} \; \D M_{B_d}, \; a_{\psi K_S}:
 C_1^{Wtt} \rightarrow  C_1^{Wtt} (1+f) ~.
\label{fmfv}
\eeq
The parameter $f$ is positive definite and real, implying that there are
no new phases in any of the quantities specified above. The size of $f$
depends on the parameters of the supersymmetric models
and the model itself \cite{nihei,goto1,goto2,baek-ko}. Given a value of
$f$, the CKM unitarity fits can be performed in these models much
the same way as they are  done for the SM. Qualitatively, the CKM-fits
in MFV models yield the following pattern for the three inner angles of
the UT:
\beq
\beta^{\rm MFV} \simeq \beta^{\rm SM}~; \; ~~\gamma^{\rm MFV} <
\gamma^{\rm SM}~; \; ~~\alpha^{\rm MFV} > \alpha^{\rm SM}~.
\label{mvfcp} 
\eeq
For example, a recent CKM-fit along these lines yields the following
central values for the three angles \cite{mfv}:
\bea
f &=& 0 ~({\rm SM}): ~~(\alpha, \beta, \gamma)_{\rm central} = (95^\circ,
22^\circ,63^\circ)~, \nn\\
f &=& 0.4 ~({\rm MFV}): ~(\alpha, \beta, \gamma)_{\rm central} =
(112^\circ, 20^\circ,48^\circ)~.
\label{mfvsmcent}
\eea
leading to $(\sin 2\beta)^{\rm SM}_{\rm central} \simeq 0.70$ and 
$(\sin 2\beta)^{\rm MFV}_{\rm central} \simeq 0.64$.  
Thus, what concerns $\sin 2 \beta$, the SM and the MFV models
give similar results from the UT-fits, unless much larger values for the
parameter $f$ are admitted which, as argued in  
Refs. \cite{nihei,goto1,goto2,baek-ko} and in this paper, is unlikely
due to the existing constraints on the MFV-SUSY parameters.
 
However, in a  general extension of the SM, one expects that
all the quantities appearing on the l.h.s. in Eq.~(\ref{fmfv})
will receive independent additional contributions. In this case, the 
magnitude and the phase of the off-diagonal elements in the $B_d^0$
-$\overline{B_d^0}$ and $B_s^0$ - $\overline{B_s^0}$ mass matrices 
can be parametrized as follows \cite{nelson,wolf-silva}:
\bea
M_{12}(B_d) &=& {\langle \bar B_d | H_{eff}^{\D
B=2} |B_d \rangle \over 2 M_{B_d}} = r_d^2 e^{2 i \theta_d}
M_{12}^{SM}(B_d)~,
\nn \\
M_{12}(B_s) &=& {\langle \bar B_s | H_{eff}^{\D
B=2} |B_s \rangle \over 2 M_{B_s}} = r_s^2 e^{2 i  \theta_s}
M_{12}^{SM}(B_s)~. 
\label{mbdgen}
\eea
where $r_d$ ($r_s$) and $\theta_d$ ( $\theta_d$) characterize,
respectively, the magnitude and the phase of the new physics
contribution to the mass difference $\D M_{B_d}$ ($\D M_{B_s}$). It
follows that a measurement of $a_{\psi K_s}$ would not measure $\sin 2
\beta$, but rather a combination $\sin 2 (\beta +
\theta_d)$. Likewise, a measurement of the CP asymmetry in the decays
$B_d^0 \to \pi \pi$ and its charge conjugate, $a_{\pi \pi}$, (assuming
that the penguin contributions are known) would not measure $\sin 2
\alpha$, but rather $\sin 2(\alpha -\theta_d)$. Very much along the
same lines, the decay $B_s \to J/\psi \phi$ and its charge conjugate
would yield a CP asymmetry $a_{\psi\phi} \simeq - \sin (\delta -
\theta_s)$, where $\delta \simeq 1/2\lambda^2 \simeq 0.02$ in the SM,
and $\lambda \simeq 0.22$ is one of the four CKM parameters in the
Wolfenstein representation \cite{wolfenstein}. Thus, the phase
$\theta_s$ could enhance the CP-asymmetry $a_{\psi\phi}$ bringing it
within reach of the LHC-experiments \cite{ball-lhc}. In this scenario,
one also expects new contributions in $M_{12}(K^0)$, bringing in their
wake additional parameters ($r_\epsilon$, $\theta_\epsilon$). They
will alter the profile of CP-violation in the decays of the neutral
kaons. In fact, sizable contributions from the supersymmetric sector
have been entertained in the literature, though it appears now
unlikely that $\epsilon_K$ and/or $\epsilon^\prime_K/\epsilon_K$
(which is a measure of direct $CP$ violation in the neutral Kaon
decays) are saturated by supersymmetry \cite{oscar,bjkp}.

It is obvious that in such a general theoretical scenario, which
introduces six {\it a priori} independent parameters, the predictive
power vested in the CKM-UT analysis is lost.  We would like to retain
this predictivity, at least partially, and entertain a theoretical
scenario which accommodates the current measurement in flavour physics,
including the recent measurements of 
$a_{\psi K_S}$, but admits additional flavour structure. A model which
incorporates these features is
introduced and discussed in section 2, using the language of minimal
insertion approximation (MIA) \cite{mia} in a supersymmetric context.
In this framework, gluinos are assumed heavy and hence have no
measurable consequences for low energy phenomenology. All FC
transitions which are not generated by the CKM mixing matrix are
proportional to the properly normalized off--diagonal elements of the
squark mass matrices:
\beq
(\d_{ij})^{U,D}_{AB} \equiv {(M^2_{ij})^{U,D}_{AB} \over
M_{\tilde q_i} M_{\tilde q_j}}
\label{miadef}
\eeq
where $i,j=1,2,3$ and $A,B=L,R$.
We give arguments why we expect that the dominant effect of the
non-CKM structure contained in the MIA-parameters is expected to
influence mainly the $b \to d$ and $s\to d $ transitions while the $b
\to s$ transition is governed by the MFV-SUSY and the SM contributions
alone.  For what concerns the quantities entering in the UT analysis,
the following pattern for the supersymmetric contributions emerges in
this model:
\bea
\label{f}
\D M_{B_s}: & C_1^{Wtt} \rightarrow & C_1^{Wtt} (1+f) \\
\e_K, \; \D M_{B_d}, \; a_{\psi K_S}: & C_1^{Wtt} \rightarrow &
         C_1^{Wtt} (1+f)+C_1^{MI} \equiv
         C_1^{Wtt} \left(1+f+ g \right)
\label{g}
\eea
where the parameters $f$ and $g=g_R + i g_I$ represent normalized
(w.r.t the SM top quark $Wtt$) contributions from the MFV and MIA
sectors, respectively.  Thus, in the UT-analysis the contribution from
the supersymmetric sector can be parametrized by two real parameters
$f$ and $g_R$ and a parameter $g_I$, generating a phase $\theta_d$,
which is in general non-zero due to the complex nature of the
appropriate mass insertion parameter.  We constrain these parameters,
taking into account all direct and indirect bounds on the
supersymmetric parameters, including the measured rates for $b \to s
\gamma$ decay \cite{cleo,belle,aleph}, $(g-2)_\mu$ from the Brookhaven
experiment \cite{gm2exp}, and the present bound on the $b \to d
\gamma$ transition, following from the experimental bound on the ratio
of the branching ratios $R(\rho \gamma/K^* \gamma) = 2{\cal B}(B^0 \to
\rho^0 \gamma)/{\cal B}(B^0 \to K^{* 0} \gamma)$ {\cite{bellebdg}.  We
do not include the quantity $\epsilon_K'/\epsilon_K$ in our analysis,
despite its impeccable measurement by the NA48~\cite{wahl} and
KTeV~\cite{glazov} Collaborations, yielding the present world average
$\Re \epsilon_K'/\epsilon_K = (1.7 \pm 0.2 ) \times
10^{-3}$, due to the inherent non--perturbative
uncertainties which have greatly reduced the impact of the
$\epsilon_K'/\epsilon_K$ measurement on the CKM phenomenology (see,
for a recent review, Ref.~\cite{buras-ckm}).

 This model, called henceforth the Extended-MFV model, leads to a
number of testable consequences, some of which are common with the
more general scenarios discussed earlier in the context of
Eq.~(\ref{mbdgen}) \cite{nelson,wolf-silva}.  Thus, for a certain
range of the argument of the MIA parameter, this model yields $a_{\psi
K_s} < a_{\psi K_s}^{\rm SM/MFV}$. For other choices of the model
parameters, this model yields a higher value for this CP asymmetry.  A
precise measurement of $a_{\psi K_s}$ would fix this argument
($=\theta_d$) and we show its allowed range suggested by the current
data. Likewise, the CP-asymmetry $a_{\pi \pi} =\sin 2(\alpha -
\theta_d)$ will be shifted from its SM-value, determined by
$\theta_d$. All $b \to d$ transitions (leading to the decays such as
$b \to d \gamma$, $b \to d \ell^+ \ell^-$, $B_d^0 \to \ell^+ \ell^-$,
where $\ell^\pm =e^\pm,\mu^\pm,\tau^\pm$, and the ratio of the mass
differences $\D M_{B_s}/\D M_{B_d}$) may turn out to be significantly
different from their SM and MFV counterparts. To illustrate this, we
work out in detail the implications for the exclusive decays $B \to
\rho^0 \gamma$ and $B^\pm \to \rho^\pm \gamma$, concentrating on the
(theoretically more reliable) ratios $R(\rho \gamma/K^* \gamma)\equiv
2 \branch (B^0\to \rho^0 \g)/ \branch (B^0\to K^* \g)$, the isospin
violating ratio $\Delta^{\pm 0}\equiv {\cal B}(B^\pm \to \rho^\pm
\gamma)/2{\cal B}(B^0 \to \rho^0 \gamma) -1$, and direct CP-asymmetry
in the decay rates for $B^- \to \rho^- \gamma$ and its charge
conjugate.  We also find that the fits of the CKM unitarity triangle
in the extended MFV model, characterized by Eqs.~(\ref{f}) and
(\ref{g}) above, admit both $\bar\rho>0$ and $\bar\rho<0$ solutions,
where $\bar\rho$ is one of the Wolfenstein
parameters~\cite{wolfenstein}. We illustrate this by working out the
predicted values of $\g$ (and $\a$) and $\D M_{B_s}$ in this model for
some specific choice of the parameters. This is in contrast with the
corresponding fits in the SM and the MFV--MSSM models, which currently
yield $\bar\rho>0$ at 2 standard deviations in the SM, with the
significance increasing in the MFV--MSSM models. The allowed CKM--fits
in the extended--MFV model with $\bar\rho<0$ imply in turn
$\gamma>\pi/2$. We note that such large values of $\gamma$ are hinted
by phenomenological analyses~\cite{hou,beneke} of the current
measurements of the branching ratios for the decays $B\to K\pi,\;
\pi\pi$~\cite{cleokp,babarkp,bellekp}. However, this inference is not
yet convincing due to the present precision of data and lack of a
reliable estimate of non--perturbative final state interactions in
these decays.

 This paper is organized as follows: In section 2, we give the outline
of our extended-MFV model.The supersymmetric contributions to the
quantities of interest ($\e_K$, $\D M_{B_s}$, $\D M_{B_d}$, $a_{\psi
K_S}$) and $R(\rho \gamma/K^*\gamma)$ are discussed in section 3,
where we also discuss the impact of the $(g-2)_\mu$ experiment on our
analysis. Numerical analysis of the parameters $(f,\vert g \vert)$,
taking into account the experimental constraints from the $b \to s
\gamma$, $b \to d \gamma$ and $(g-2)_\mu$, is presented in section
4. A comparative analysis of the unitarity triangle in the SM, MFV and
the Extended-MFV models is described in section 5, where we also show
the resulting constraints on the parameters $g_R$ and $g_I$ and the
CP-asymmetry $a_{\psi K_s}$. The impact of the Extended-MFV model on
the $b \to d \gamma$ transitions are worked out in section 6. Section
7 contains a summary and some concluding remarks. The explicit stop
and chargino mass matrices are displayed in Appendix A and some loop
functions encountered in the supersymmetric contributions are given in
Appendix B.

\section{Outline of the model}
\label{outline}
The supersymmetric model that we consider is a generalization of the
model proposed in Ref.~\cite{giudice}, based on the assumptions of
Minimal Flavour Violation with heavy squarks (of the first two
generations) and gluinos.  The
charged Higgs and the lightest chargino and stop masses are required
to be heavier than $100 \; \gev$ in order to satisfy the
lower bounds from direct searches.
 The rest of the SUSY spectrum is assumed to be almost
degenerate and heavier than $1 \; \tev$. In this framework the
lightest stop is almost right--handed and the stop mixing angle (which
parameterizes the amount of the left-handed stop $\tilde t_L$ present in
the lighter mass
eigenstate) turns out to be of order $O(M_W / M_{\tilde q}) \simeq
10\%$; for definiteness we will take $|\theta_{\tilde t}| \leq \p
/10$.

The assumption of a heavy ($\ge 1$ TeV) gluino totally suppresses any
possible gluino--mediated SUSY contribution to low energy
observables. On the other hand, the presence of only a single light
squark mass eigenstate (out of twelve) has strong consequences due to
the rich flavour structure which emerges from the squark mass
matrices.  As discussed in the preceding section, adopting the
MIA-framework~\cite{mia}, all the FC effects which are not generated
by the CKM mixing matrix are proportional to the properly normalized
off--diagonal elements of the squark mass matrices (see
Eq.~(\ref{miadef})).  In order to take into account the effect of a
light stop, we exactly diagonalize the $2\times 2$ stop system and
adopt the slightly different MIA implementation proposed in
Ref.~\cite{mia2}.  In this approach, a diagram can contribute sizably
only if the inserted mass insertions involve the light stop. All other
diagrams require necessarily a loop with at least two heavy ($\geq 1
\; \tev$) squarks and are therefore automatically suppressed.
This leaves us with only two unsuppressed flavour changing sources
other than the CKM matrix, namely the mixings $\tilde u_L - \tilde
t_2$ (denoted by $\d_{\tilde u_L \tilde t_2}$) and $\tilde c_L -
\tilde t_2$ (denoted by $\d_{\tilde c_L \tilde t_2}$).  We note
that $\d_{\tilde u_L \tilde t_2}$ and $\d_{\tilde c_L \tilde t_2}$ are
mass insertions extracted from the up--squarks mass matrix after the
diagonalization of the stop system and are therefore linear
combinations of $(\d_{13})^U_{LR}$, $(\d_{13})^U_{LL}$ and of
$(\d_{23})^U_{LR}$, $(\d_{23})^U_{LL}$, respectively.

Finally, a comment on the normalization that we adopt for the mass
insertions is in order. In Ref.~\cite{casas} it has been pointed out
that $(\d_{i3})^U_{LR}$ must satisfy an upper bound of order $2 m_t/
M_{\tilde q_i}$ in order to avoid charge and colour breaking minima
and directions unbounded from below in the scalar potential. We
normalize the insertions relevant to our discussion so that, in the
limit of light stop, they automatically satisfy this constraint: 
\beq
\label{delta}
\d_{\tilde u(\tilde c)_L \tilde t_2} \equiv {M^2_{\tilde u(\tilde c)_L \tilde
t_2}
\over M_{\tilde t_2} M_{\tilde q}} { |V_{td(s)}| \over V_{td(s)}^*} \;
.  \eeq 
This definition includes the phase of the CKM element $V_{td(s)}$.  In
this way, deviations from the SM predictions, for what concerns CP
violating observables, will be mainly associated with complex values
of the mass insertion parameters. For instance, as we will argue in
the following, the CP asymmetry $a_{\psi K_S}$ in the decay $B
\rightarrow J/\psi K_s$ can differ from the SM expectation only if
$\arg \d_{\tilde u_L \tilde t_2} \neq 0$. In general, the two phases
must not be aligned with the respective SM-phases entering in the
box diagram:
\beq
\arg {M^2_{\tilde u(\tilde c)_L \tilde t_2} \over M_{\tilde t_2} M_{\tilde q}}
\neq \arg V_{td(s)}^* \; .
\eeq

The insertion $\d_{\tilde c_L \tilde t_2}$ characterizes the
$b\rightarrow s$ transitions and it enters in the determination of the
$B_s - \bar B_s$ mass difference, the $b\rightarrow s \gamma$ decay
rate, and observables related to other FCNC decays such as $b \to s
\ell^+ \ell^-$. For what concerns the $ b \to s \gamma$ decay,
previous analyses~\cite{lunghi} pointed out that contributions
proportional to this insertion can be as large as the SM one. The
experimental results for the inclusive branching fraction are
\beq
\branch (B\rightarrow X_s \gamma) = \cases{ 
(3.19 \pm 0.43_{\scriptscriptstyle stat} \pm 0.27_{\scriptscriptstyle syst}) 
      \times 10^{-4} & CLEO~\cite{cleo} \cr
(3.11 \pm 0.80_{\scriptscriptstyle stat} \pm 0.72_{\scriptscriptstyle syst}) 
      \times 10^{-4} & ALEPH~\cite{aleph} \cr
(3.36 \pm 0.53_{\scriptscriptstyle stat} \pm 42_{\scriptscriptstyle syst} 
      \pm ({}^{0.50}_{0.54})_{\scriptscriptstyle model}) 
      \times 10^{-4} & BELLE~\cite{belle}. \cr
}
\eeq
Combining these results and adding the errors in quadrature we obtain
the following world average for the inclusive branching ratio
\beq
\label{bsgM} 
\branch (B\rightarrow X_s \gamma) = (3.22 \pm 0.40) \times 10^{-4}
\eeq
yielding the following $\cl{95}$ experimentally allowed range
\beq
\label{bsg} 
2.41 \times 10^{-4} \leq 
\branch (B\rightarrow X_s \gamma) 
\leq 4.02 \times 10^{-4} \; . 
\eeq
Using the LO theoretical expression for this branching ratio, the
following bound is obtained:
\beq 
\label{c7lo}
0.30 \leq 
\left| C_7^{eff,LO} (m_b)\right|
\leq 0.40 
\eeq
where $C_7^{eff,LO}(m_b)$ is the relevant Wilson coefficient evaluated
in the LL approximation and has the value $-0.316$ in the SM. Analyses
of the NLO SM~\cite{bsgnlo} and
SUSY~\cite{2hdm,giudice,giudice1,carena} contributions (for the latter
only a limited class of SUSY models were considered) showed that the
LO result can receive substantial corrections.  Notice that the NLO
analysis presented in Refs.~\cite{2hdm,giudice,giudice1,carena} is
valid exactly for the class of models that we consider over all the
SUSY parameter space (including the large $\tan \beta$
region). Implementing their formulae and allowing the SUSY input
parameters to vary within the range that we will further discuss in
section~\ref{susyanalysis}, we find that, up to few percents, the
branching ratio is given by
\bea 
\label{bsgnlo}
\branch (B\rightarrow X_s \gamma) &=& 
22.23 \; \left[ C_7^{eff,NLO} (m_b) -0.061 \right]^2 +0.264 \; ,\\
C_7^{eff,NLO} (m_b) &=& 
-0.175 + 0.666 \; C_7^{eff,NLO} (M_W) + 0.093 \; C_8^{eff,NLO} (M_W) \; .
\eea
The explicit expressions for $C_{7,8}^{eff,NLO} (M_W)$ can be found in
Refs.~\cite{2hdm,giudice,giudice1,carena}. Combining \eq{bsgnlo} and the
bound (\ref{bsg}), we obtain 
\bea 
-0.35 \leq C_7^{eff,NLO} (m_b) \leq -0.24 
\quad\quad \hbox{or} \quad \quad 
0.36 \leq C_7^{eff,NLO} (m_b) \leq 0.49 \; . 
 \eea
Notice that in Refs.~\cite{giudice1,carena} it is pointed out that in
order to get a result stable against variations of the heavy SUSY
particles scale, it is necessary to properly take into account all
possible SUSY contributions and to resum all the large logarithms that
arise. The inclusion of the insertion $\d_{\tilde c_L \tilde t_2}$ in
this picture, in particular, should not be limited to the LO matching
conditions but should instead extend to the complete NLO analysis.
This program is clearly beyond the scope of the present paper.
Moreover, one finds that, including the NLO corrections, the SM almost
saturates the experimental branching ratio. In view of this we choose
not to consider $\d_{\tilde c_L \tilde t_2}$ in our analysis. Of
course, the SUSY contribution from the MFV sector is still there, but
it is real relative to the SM. The assumption of neglecting
$\d_{\tilde c_L \tilde t_2}$ will be tested in CP-asymmetries ${\cal
A}_{\rm CP}(b \to s \gamma)$ and ${\cal A}_{\rm CP}(B \to K^* \gamma)$
at the B-factories.  Notice that the exclusion of $\d_{\tilde c_L
\tilde t_2}$ from our analysis introduces strong correlations between
the physics that governs $b\ra d$ and $b\ra s$ transitions, such as
the ratio $\Delta M_{B_s}/\Delta M_{B_d}$, which would deviate from
its SM (and MFV model) values.

The free parameters of the model are the common mass of the heavy
squarks and gluino ($M_{\tilde q}$), the mass of the lightest stop
($M_{\tilde t_2}$), the stop mixing angle ($\theta_{\tilde t}$), the
ratio of the two Higgs vevs ($\tan \b_S$~\footnote[2]{We adopt the
notation $\b_S$ in order not to generate confusion with the inner
angle of the unitarity triangle which is denoted by $\b$}), the two
parameters of the chargino mass matrix ($\m$ and $M_2$), the charged
Higgs mass ($M_{H^\pm}$) and $\d_{\tilde u_L \tilde t_2}$.  All these
parameters are assumed to be real with the only exception of the mass
insertion whose phase in not restricted {\it a priori}. In this way we
avoid
any possible problem with too large contributions to flavour
conserving CP violating observables like the electric dipole moments of
the leptons, hadrons and atoms.

In the next section we analyze the structure of the SUSY contributions 
to the observables related to the determination of the unitarity triangle,
namely $\e_K$, $\D M_{d,s}$ and $a_{\psi K_S}$. 
 
\section{SUSY contributions}
\label{susy}
The effective Hamiltonian that describes $\D F=2$ transitions can be
written as 
\begin{eqnarray}
    {\cal{H}}_{eff}^{\Delta F=2}
&=&
    -\frac{G_{F}^{2} M_{W}^{2}}{(2 \pi)^{2}} (V_{tq_1} V_{tq_2}^*)^{2}\ 
    \left( C_{1}(\mu)\  \bar{q}^{\alpha}_{2L}\gamma^{\mu}
    q^{\alpha}_{1L}\cdot \bar{q}^{\beta}_{2L}\gamma_{\mu}q^{\beta}_{1L}\ 
    +\  C_{2}(\mu)\  \bar{q}^{\alpha}_{2L}q^{\alpha}_{1R}\cdot 
    \bar{q}^{\beta}_{2L} q^{\beta}_{1R}
       \right. \nn \\
& &
       \left. 
    +\ C_3(\mu)\ \bar{q}^{\alpha}_{2L}q^{\beta}_{1R}\cdot 
    \bar{q}^{\beta}_{2L}q^{\alpha}_{1R}\right) + {\it h. c.}\; ,
\end{eqnarray}
where $\alpha, \beta$ are colour indices and $(q_1,q_2)$ = $(s,d)$,
$(b,d)$, $(b,s)$ for the $K$, $B_d$ and $B_s$ systems respectively.

In this framework, as previously explained, gluino contributions are
negligible; therefore, we have to deal only with charged Higgs (which
obeys the SM CKM structure) and chargino mediated box diagrams. Let us
comment on the latter. The dominant graphs must involve exclusively
the lightest stop eigenstate since the presence of an heavy ($\geq 1\;
\tev$) sparticle would definitely suppress their contribution.
Moreover, Feynman diagrams that contribute to $C_3$ are substantially
suppressed with respect to diagrams that contributes to $C_1$.  In
fact, for what concerns the $B$ system, the vertices $b_R - \tilde H -
\tilde t_2$ and $b_L - \tilde H - \tilde t_2$ are proportional,
respectively, to $m_b\sin \theta_{\tilde t}/(\sqrt{2} M_W \cos \b_S) $
and $m_t\cos \theta_{\tilde t}/(\sqrt{2} M_W \sin \b_S) $. Their ratio
is thus of order $(m_b/m_t \tan\b_S \tan \theta_{ \tilde t})^2$ which,
even in the large $\tan \b_S$ regime, is damped and not much larger
than $O(0.1)$. In the K system $m_b$ must be replaced by $m_s$ and the
suppression is even stronger. Notice that in frameworks in which the
split between the two stop mass eigenstates is not so marked, this
argument fails. Diagrams mediated by the exchange of both stops must
be considered and it is possible to find regions of the parameter
space (for large $\tan \b_S$) in which SUSY contributions to $C_3$ are
indeed dominant~\cite{oscar}.  We note that, the $\tan^4 \b_S$
enhanced neutral Higgs contributions to the coefficients $C_{2,3}$,
whose presence is pointed out in Ref.~\cite{janusz}, do not impact
significantly for the range of SUSY parameters that we consider ($\tan
\b_S< 35$ and $|\theta_{\tilde t}|< \pi/10$).

The total contribution to $C_1$ can thus be written as
\beq
C_1^{tot} (M_W) = C_1^W (M_W)+ C_1^{H^\pm} (M_W)+ C_1^{\chi} (M_W)+ C_1^{MI} (M_W)\ .
\eeq
The explicit expressions for the various terms are:
\bea
\label{c1w}
C_1^W &=& C_1^{Wtt} + \left(V_{cq_1}V_{cq_2}^*\over  
          V_{tq_1} V_{t q_2}^* \right)^2 C_1^{Wcc} + 2 \; {
          V_{cq_1} V_{cq_2}^*\over  
          V_{tq_1} V_{tq_2}^*} \; C_1^{Wtc} \; ,\\
C_1^{H^\pm} &=& {x_t^2 \over 4\ \tan^4 \b_S}\ Y_1(x_H,x_H,x_t,x_t)+ 
            {x_t^2  \over 2\ \tan^2 \b_S}\  Y_1(1,x_H,x_t,x_t) \nn\\
            && - {2 x_t \over \tan^2 \b_S}  \ Y_2(1,x_H,x_t,x_t) \ , \\
C_1^{\chi} &=& \sum_{i,j=1}^2 \left| \tilde  V_{i2} {m_t\cos
\theta_{\tilde t}\over \sqrt{2} 
               M_W \sin \b_S} - \sin \theta_{\tilde t} \tilde  V_{i1}^* \right|^2
               \left| \tilde V_{j2} {m_t\cos \theta_{\tilde t} \over
\sqrt{2} 
               M_W \sin \b_S} - \sin \theta_{\tilde t} \tilde  V_{j1}^* \right|^2 \nn\\
           & & \times Y_1(x_{\tilde t},x_{\tilde t},x_{\chi_i},x_{\chi_j})  \ , \\
\label{c1mi}
C_1^{MI} &=& \left| V_{ud}\over V_{td} \right|^2 \sum_{i,j=1}^2  \tilde V_{i1}  \tilde V_{j1} \left(
              \tilde V_{i2} {m_t\cos \theta_{\tilde t}\over \sqrt{2} M_W
\sin \b_S} - \sin
             \theta_{\tilde t}  \tilde V_{i1}^* \right)^* \left( \tilde V_{j2}
 {m_t\cos
             \theta_{\tilde t} \over \sqrt{2} M_W \sin \b_S} - \sin \theta_{\tilde t}
              \tilde V_{j1}^* \right)^* \nn\\ &&\times Y_1^{MI}(x_{\tilde q},x_{\tilde
             t},x_{\chi_i},x_{\chi_j}) \; \d_{\tilde u_L \tilde t_2}^2 \nn\\
         &\equiv& \overline C_1^{MI} \; \d_{\tilde u_L \tilde t_2}^2
\eea
where $C_1^{W\a\b} = G (x_\a,x_\b)$ and $x_\a = m_\a^2/M_W^2$. The
contributions to $C_1^\chi$ and $C_1^{MI}$ come from the Feynman
diagrams shown in \fig{diag}. The conventions we adopt for the
chargino mass matrix, the exact definition of the stop mixing angle
and the explicit expressions for the loop functions $G$, $Y_1$, $Y_2$
and $Y_1^{MI}$ can be found in the appendices. Notice that, if we
restrict to the $B_d$ and $B_s$ cases, one may neglect the
terms $C_1^{Wtc}$ and $C_1^{Wcc}$ in \eq{c1w} as they are
suppressed by small CKM factors. In the $K$ system, on the other hand,
it is necessary to consider all the terms.

\eq{c1mi} describes the impact of a non--zero mass insertion on the
$B_d$ {\em and} $K$ systems. The corresponding contribution to the $B_s$
system is obtained via the substitution 
\beq 
\left|V_{ud}\over V_{td} \right|^2 \rightarrow
\left|V_{us}\over V_{ts} \right|^2 \; .  
\eeq
This implies that the impact of this diagram on the $B_s$ system is
reduced by a factor $|V_{us} V_{td}/ V_{ts}|^2 \simeq 0.0016$ with
respect to the $B_d$ and $K$ systems.  Since, as we will show in the
forthcoming analysis, $C_1^{MI}$ is not likely to exceed by more than
twice the SM contribution, it is clear that any effect in the $B_s$
system from the mass insertion is completely negligible.

As already discussed in the introduction, the following structure
of the SUSY contributions  emerges in the class of model described above: 
\bea 
\D M_{B_s}: & C_1^{Wtt} \rightarrow & C_1^{Wtt} (1+f) \nn \\ 
\e_K, \; \D M_{B_d}, \; a_{\psi K_S}: & C_1^{Wtt} \rightarrow &
         C_1^{Wtt} (1+f)+C_1^{MI} \equiv 
         C_1^{Wtt} \left(1+f+ g \right) \nn 
\eea 
where the parameters $f$ and $g$ represent normalized 
contributions from the MFV and MIA sectors, respectively, 
\bea
f &\equiv& {(C_1^{H^\pm} + C_1^\chi) / C_1^{Wtt}} \\
g &\equiv& g_R + i g_I \equiv {\overline C_1^{MI} \d_{\tilde u_L \tilde t_2}^2 / C_1^{Wtt}} \ .
\eea
The impact of the SUSY models on the observables we are interested in
is then parametrized by three real parameters $f$, $g_R$ and $g_I$.
We recall here that in the limit $g\to 0$, the above parametrization
reduces to the one given in Ref.~\cite{ali1,ali2} for the minimal
flavour violation and the MSSM cases (More generally, for all models
in which the CKM matrix is the only flavour--changing structure).  The
absence of any CKM phase in \eq{c1mi} as well as in the definition of
$g$ reflects the definition of the mass insertion given in \eq{delta}.

Note that $f$ and $g$ are functions of the SUSY parameters that enter
in the computation of many other observables that are not directly
related to CP violation. This implies that it is possible to look for
processes that depend on the same SUSY inputs $f$ and $g$. For
example, the presence of non trivial experimental bounds on the
$\vert \Delta B=2\vert,~\Delta Q=0$ transitions can induce interesting
correlations with the UT analysis. Likewise, the inclusive radiative decay
$B\ra X_s \g$ and $(g-2)_{\mu}$, the anomalous magnetic moment of the
muon, are susceptible to supersymmetric contributions. On the
other hand, it is necessary to search for observables that can put
constraints on the insertion $ \d_{\tilde u_L \tilde t_2}$. In this
context, the transitions $b\ra d \g$, $B^0\ra \rho^0 \g$ and $B^\pm
\ra \rho^\pm \g$ are obvious places to look for $\d_{\tilde u_L \tilde
t_2}$--related effects.

Concerning now the $(g-2)_{\mu}$ constraint, we recall that 
the Brookhaven Muon $(g-2)$-collaboration has recently measured with
improved precision the anomalous magnetic moment of the positive
muon. The present world average for this quantity is~\cite{gm2exp}
\beq 
a_{\mu^+}(\hbox{exp}) =
116592023(15) \times 10^{-10} \ .  
\eeq
The contribution to $a_{\mu^+}$ in the SM arises from the QED and
electroweak corrections and from the hadronic contribution which
includes both the vacuum polarization and light by light
scattering~\cite{marciano12}. The error in the SM estimate is
dominated by the hadronic contribution to $a_{\mu^+}$ and is obtained
from $\sigma (e^+ e^- \rightarrow \mbox{hadrons})$ via a dispersion
relation and perturbative QCD. The light-by-light hadronic contribution
is, however, completely theory-driven. Several competing estimates of 
$a_{\mu^+}^{had}$ exist in the literature, reviewed recently in
Ref.~\cite{MR01}. We briefly discuss a couple of representative estimates
here.  

In order to minimize the experimental errors in the low-$s$ region, Davier
and H\"{o}cker supplemented the $e^+ e^- \rightarrow \pi^+ \pi^-$
cross section using data from tau decays. Using isospin symmetry
they estimate~\cite{davier}
\beq
a_{\mu^+}^{had} = (692.4 \pm 6.2) \times 10^{-10} \; .
\label{dh98}
\eeq
An updated value of $a_{\mu^+}^{had}$ using earlier estimates of
Eidelman and Jegerlehner~\cite{jeger01} supplemented by the
recent data from CMD and BES collaborations yields~\cite{jeger2}
\beq
a_{\mu^+}^{had} = (698.75 \pm 11.11) \times 10^{-10} \; .
\label{ej95}
\eeq
Alternatively, calculating the Adler function from $e^+ e^-$ data and
perturbative QCD for the tail (above $11 \; \gev$) and calculating the
shift in the electromagnetic fine structure constant $\D \a^{had}$ in
the Euclidean region, Jegerlehner quotes~\cite{jeger2}
\beq
a_{\mu^+}^{had} = (697.4 \pm 10.45) \times 10^{-10} \; .
\label{jege}
\eeq
These estimates are quite compatible with each other, though the
errors in Eqs.~(\ref{ej95}) and (\ref{jege}) are larger than in
Eq.~(\ref{dh98}). Adopting the theoretical estimate from Davier and
H\"ocker~\cite{davier}, one gets
\beq 
a_{\mu^+}(\hbox{SM}) =
11659163(7) \times 10^{-10} \ ,  
\eeq
yielding
\beq
\delta a_{\mu^+} \equiv a_{\mu^+}(\hbox{exp}) - a_{\mu^+}(\hbox{SM}) 
 = +43(16) \times 10^{-10},
\label{deltaamu}
\eeq
which is a $2.6 \; \sigma$ deviation from the SM.  Using, however,
the estimate from Jegerlehner in \eq{jege}, gives
\begin{equation}
\delta a_{\mu^+} = +37(18) \times 10^{-10},
\label{deltaamu1}
\end{equation}
which amounts to about $2\; \s$ deviation from the SM. Thus, on the
face-value, there exists a 2 to 2.6 $\sigma$ discrepancy between the
current experiments on $(g-2)_\mu$ and SM-estimates.

In SUSY theories, $a_{\mu^+}$ receives contributions via vertex
diagrams with $\chi^0$--$\tilde \mu$ and $\chi^\pm$--$\tilde \nu$
loops~\cite{marciano12,moroi,nath12,everett,gm2pap,gondolo,
nath3,moroi2,ellisgm2,martin,ellisgm21,baer}.  The chargino diagram
strongly dominates over almost all the parameter space.  The chargino
contribution is~\cite{moroi} (see also Ref.~\cite{nath12} for a
discussion on $CP$ violating phases):
\begin{equation}
\delta a_{\mu^+}^{\chi \tilde\nu} = {g_2^2\over 8 \pi^2}
{M_\mu^2\over M_{\tilde\nu}^2}
 \sum_{i=1}^2 \left\{
\left[ {M_\mu^2 \; \Re (\tilde U_{i2}^2) \over 2 M_W^2 \cos^2\b_S}  + \Re (\tilde V_{i1}^2)
\right] F_1 (x_i)
 -\frac{M_{\chi_i} \hbox{Re} (\tilde U_{i2} \tilde V_{i1})}
      {\sqrt{2} M_W \cos \beta}
 F_3 (x_i) \right\} \; ,
\label{gm2char}
\end{equation}
where $x_i = M_{\chi_i}^2/M_{\tilde\nu}^2$ and the loop functions
$F_1$ and $F_3$ are given in the appendices. \eq{gm2char} is dominated
by the last term in curly brackets whose sign is determined by
$\hbox{sign}[\hbox{Re}(\tilde U_{12} \tilde V_{11}) ] = - \hbox{sign}
[\hbox{Re}(\mu)]$ (note that we have $M_{\chi_1}<M_{\chi_2}$).  Taking
into account that the Brookhaven experiment implies $\delta a_{\mu^+}
> 0 $ at 2 to 2.6 $\s$, it is clear that the $\mu>0$ region is currently
favoured.

Finally let us focus on $b\ra d \g$ decays.  To the best of our
knowledge, there is no direct limit on the inclusive decay $b\ra d
\gamma$.  The present experimental upper limits on some of the
exclusive branching ratios are
\bea 
&&\branch (B^0 \ra \r^0 \g) < 0.56 \times 10^{-5} \;\;\; (\cl{90}) 
  \mbox{\cite{bellebdg}} \; , \\
&&\branch (B^+ \ra \r^+ \g) < 1.3 \times 10^{-5} \;\;\; (\cl{90}) 
  \mbox{\cite{cleobdg}} \; , \\ 
&& R(\r \g/K^* \g)\equiv \branch (B \ra \r \g) / \branch (B \ra K^* \g) <
0.28 \;\;\; (\cl{90}) 
    \mbox{\cite{bellebdg}} \; .  
\eea 
In the numerical analysis we will use only the last constraint since
the ratio of branching ratios is theoretically cleaner as only the ratio
of the form factors is involved, which is calculable more reliably.
Concentrating on the neutral $B$-decays, the LO
expression for $R(\r \g/K^* \g)$
is~\cite{aliff,alibdg}
\beq 
R(\r \g/K^* \g) = { 2 \branch
(B^0 \ra \r^0(770) \g) \over \branch (B \ra K^* \g)} = \left| V_{td}
\over V_{ts} \right|^2 \left( M_B^2-M_\r^2 \over M_B^2-M_{K^*}^2
\right)^3 \xi \left| C_7^d (m_b) \over C_7^s (m_b)
\right|^2 \; ,
\label{ratio}
\eeq 
where 
\beq
\xi = \left| F_1^{B^0\ra \rho^0} \over F_1^{B^0\ra K^{*0}} \right|^2 \; ,
\eeq
with $F_1^{B^0\ra \rho^0,K^{*0}}$ being the form factors involving the
magnetic moment (short--distance) transition; $M_B=5.2794 \; \gev$,
$M_{K^*} = 0.8917 \; \gev$, $M_{\r^0} = 0.7693 \; \gev$ and $\xi =
0.58$~\cite{aliff}. $C_7^{(d,s)}(m_b)$ is the Wilson coefficient of
the magnetic moment operator for the transition $b\ra (d,s)$ computed
in the leading order approximation.  Note that the annihilation
contribution (which is estimated at about $25 \; \%$ in $B^\pm\ra \rho^\pm
\g$~\cite{wyler,ab95}) is suppressed due to the unfavourable
colour factor and the electric charge of the d--quark in $B^0$, and
ignored here. In the SM, these two Wilson coefficients coincide while,
in the SUSY model we consider, they differ because of the effect of
the insertion $\d_{\tilde u_L \tilde t_2}$:
\bea
C_7^s (M_W) &=& C_7^W (M_W) + C_7^{H^\pm} (M_W) + C_7^\chi (M_W) \; , \\
C_7^d (M_W) &=& C_7^s (M_W) + C_7^{MI} (M_W) \; .
\label{c7d}
\eea
where $C_7^W$, $C_7^{H^\pm}$ and $C_7^\chi$ are, respectively, the SM, the
charged Higgs and the chargino contributions and their explicit
expressions can be found for instance in Refs.~\cite{2hdm,giudice}.
The explicit expression for the mass insertion contribution is
\bea
C_7^{MI} (M_W) &=& \left| V_{ud} \over V_{td} \right| \sum_{i=1}^2 
       {M_{\tilde q} M_{\tilde t_2} M_W \over 6 M_{\chi_i}^3} \tilde V_{i1} \left\{
       \left( {\tilde V_{i2}^* m_t \cos \theta_{\tilde t} \over \sqrt{2}
M_W \sin\b_S}
       - \tilde V_{i1}^* \sin \theta_{\tilde t} \right) 
       {M_W \over M_{\chi_i}} f_1^{MI}(x_{\tilde t_2 \chi_i},x_{\tilde q \chi_i}) 
             \right. \nn \\ 
         & & \left. +{\sqrt{2} \tilde U_{i2}  \sin \theta_{\tilde t}\over \cos\b_S} 
             f_2^{MI}(x_{\tilde t_2 \chi_i},x_{\tilde q \chi_i}) \right\}
             \d_{\tilde u_L \tilde t_2} \\
              &\equiv& \overline C_7^{MI} \d_{\tilde u_L \tilde t_2}
\label{c7mi}
\eea
where the loop functions $f_1^{MI}$ and $f_2^{MI}$ are given in 
Appendix B. Using Eqs.~(\ref{c7d}) and~(\ref{c7mi}) it is possible to
rewrite the ratio $R(\rho \gamma/K^* \gamma)$ in the following way:
\beq 
R (\rho \gamma/K^* \gamma)= R^{SM} \left| 1 + \d_{\tilde u_L
\tilde t_2} {\eta^{16\over 23}\overline C_7^{MI} \over C_7^s (m_b)}
\right|^2 
\label{rdelta}
\eeq 
in which $\overline C_7^{MI}$ and $C_7^s (m_b)$ are both real,
$\eta^{16\over 23}$ is a QCD factor numerically equal to 0.66 and we
have abbreviated $R (\rho \gamma/K^* \gamma)^{SM}$ by $R^{SM}$ for
ease of writing.

\section{Numerical Analysis of SUSY Contributions}
\label{susyanalysis}
In this section we study the correlations between the possible values
of the parameters $f$ and $g$ as well as the numerical impact of $b\ra
s\g$, $\d_{a_{\m^+}}$ and $b\ra d \g$ with the following experimental
constraints 
\bea
&&2.41 \times 10^{-4} \leq \branch (B\ra X_s \g) \leq 4.02 \times 10^{-4} \;\;\; (\cl{95}) \nn \\
&&10 \times 10^{-10} \leq \d a_{\mu^+} \leq 74  \times 10^{-10} \;\;\; (\cl{95}) \label{constraints}\\
&&R(\r \g/K^* \g) \leq 0.28 \;\;\; (\cl{90}) \nn
\eea
where we have used the estimates of $\d a_{\mu^+}({\rm SM})$ from
Eq.~(\ref{dh98}). We perform the numerical analysis by means of high
density scatter 
plots varying the SUSY input parameters over the following ranges:
\beq
\cases{
 \makebox[1.4cm]{$\mu$} =\;\; (100 \div 1000) \; \gev \; ,  & \cr
 \makebox[1.4cm]{$ M_2$} =\;\; (100 \div 1000) \; \gev \; ,   & \cr
 \makebox[1.4cm]{$\tan \b_S$} =\;\; 3 \div 35 \; ,   & \cr
 \makebox[1.4cm]{$M_{H^\pm}$} =\;\; (100 \div 1000) \; \gev  \; ,  & \cr
 \makebox[1.4cm]{$M_{\tilde t_2}$} =\;\; (100 \div 600) \; \gev \; ,   & \cr
 \makebox[1.4cm]{$\theta_{\tilde t}$} =\;\; -0.3 \div 0.3 \; .  & \cr}
\label{susyparams} 
\eeq 
Notice that, according to the discussion of the previous section, we
restrict the scatter plot to positive $\mu$ values only. Negative
values are strongly disfavoured both by the $\d a_{\mu^+} \geq 0$
bound and by the $b\ra s\g$ branching ratio. In fact, it is
possible to show that if $\m$ is negative, the chargino contributions
to $b\ra s\g$ tend to interfere constructively with the SM and the
charged Higgs ones. In order not to exceed the experimental upper
limit, a quite heavy SUSY spectrum is thus required. In such a
situation, high $f$ and $g$ values are quite unlikely. 

In \fig{fg} we plot the points in the $(f,|g|)$ plane that survive the
$b\ra s\g$, $\d a_{\mu^+}$ and $b\ra d\g$ constraints. Scanning over
the parameters given in \eq{susyparams} we find that the constraints
in \eq{constraints} restrict $f$ and $|g|$ to lie essentially in the
range $f<0.4$, $|g|<2.0$ . We also find that the sign of $\overline
C_1^{MI}/ C_1^{Wtt}$ is positive over all the SUSY parameter space
that we scanned.

The impact of $\d a_{\mu^+}$ on our analysis is not very strong, once
we limit the scanning to the $\m>0$ region only. Moreover, as follows
from \eq{gm2char}, the size of the chargino contribution to $\d
a_{\mu^+}$ is controlled by the mass of the muon sneutrino.  In our
framework, $m_{\tilde \n}$ is a free parameter and we have imposed the
lower bound $\d a_{\mu^+} \geq 10 \times 10^{-10}$ in the loosest
possible way by choosing $m_{\tilde \n}=100 \; \gev$ (a value that is
reasonably safe against direct search constraints).  On the other
hand, we can not reject points which give a too large contribution to
$\d a_{\mu^+}$ because a large enough sneutrino mass can always
suppress the SUSY diagram and reduce $\d a_{\mu^+}$ to a value smaller
than $74 \times 10^{-10}$. Notice that, if $m_{\tilde \n} \geq 300 \;
\gev$, all the points that we consider do satisfy the upper bound: in
order to obtain larger contributions it is necessary to impose a very
light sneutrino mass.  Only models in which the squark and the slepton
mass spectra depend on the same inputs will be able to fully exploit
the correlation between the anomalous magnetic moment of the muon and
observables related to $B$ physics.

The impact of the $b\ra d\g$ constraint is taken into account 
by imposing the following upper bound on the 
mass insertion:
\beq 
\left| \d_{\tilde u_L \tilde t_2}\right| <
\left(\sqrt{0.28 \over R^{SM}}-1\right)
 \left| C_7^s (m_b) \over \eta^{16\over
23} \overline C_7^{MI} \right| \; .
\label{dlim}
\eeq
Again, we find that with the current experimental bound 
$R(\r \g/K^* \g)< 0.28$, most of the otherwise allowed $(f,|g|)$
region survives. This situation will change with improved limits (or
measurements of $R(\r \g/K^* \g)$).

In \fig{fgp} we perform the same analysis presented in \fig{fg} but we
allow only for points that give a positive sign for the Wilson
coefficient $C_7^s$ computed in the LO approximation. The issue
whether it is possible or not to change the sign of $C_7^s$ depends on
the model and has been long debated in the literature. In particular,
this sign strongly characterizes the behaviour of the
forward--backward asymmetry and of the dilepton invariant mass in
$b\ra s\ell^+ \ell^-$ transitions, as well as the sign of the isospin
violating ratio $\Delta$ (see below) and of the CP-violating asymmetry
in the radiative decays $B \to \r \g$ \cite{alibdg}. We use $C_7^{LO}$
for calculating the CP-asymmetry in $B \to \r \g$. The quantity
$\Delta$ in the NLO approximation requires the Wilson coefficient
$C_7^{NLO}$. However, as shown in Ref.~\cite{alibdg}, $\Delta$ is
stable against NLO vertex corrections.  Recently, also the so--called
hard spectator corrections have been calculated to $O(\alpha_s)$ in
$B\to \rho \gamma$~\cite{ap01,bb01} with the result that $\Delta$ is
stable also against these corrections.  We refer to
Refs.~\cite{alibdg, cho,aliball,lunghi,lunghi3} for a comprehensive
review of the positive $C_7$ phenomenology. In \fig{fgp}, open circles
represent points that satisfy the $b\ra s\g$ and $\d a_{\mu^+}$
constraints. The black dots show what happens when the experimental
bound $R(\r \g/K^* \g)<0.28$ is imposed. In implementing this
constraint we use Eq.~(\ref{dlim}).  If for a given point
$\d^{\mbox{\tiny lim}}$ turns out to be smaller than 1, we plot
$|g|_{\d=\d^{\mbox{\tiny lim}}}$, otherwise we set $\d=1$.  It is
important to note that the dependence of $C_7^{MI}$ and $C_1^{MI}$ on
the mass insertion is, respectively, linear and quadratic. From
\fig{fgp} one sees that all the points that are compatible with a
positive $C_7$ provide, indeed, a too large contribution to $b\ra d
\g$, and hence are effectively removed by the cut on $R(\r \g/K^*
\g)$.  This result is quite reasonable because, in order to change the
sign of $C_7$, a large positive chargino contribution is needed: since
$C_7^\chi$ and $\overline C_7^{MI}$ depend on the same input
parameters we expect their magnitude to be closely correlated. In
\fig{c7c7}, we show explicitly the correlation between $C_7^s (m_b)$
and $\overline C_7^{MI} (m_b)\equiv \eta^{16\over 23} \overline
C_7^{MI} (M_W)$ in both the negative and positive $C_7^s$ allowed
regions. In the second plot, in particular, $|C_7^{MI} (m_b)|$ turns
out to be greater than one for all the points: this implies that the
mass insertion constraint is always non trivial. The strong bound
shown in \fig{fgp} is obtained by taking into account that $g$ depends
quadratically on $\d_{\tilde u_L \tilde t_2}$.  The conclusion is that
if $C_7^s>0$ is experimentally established, our analysis implies
strong constraints on the quantity $|g|$ from $b \to d \g$ decays,
permitting only small deviations from the MFV-value: $|g|=0$.

\section{Unitarity Triangle Analysis}
In section~\ref{susy} we have shown that the impact of this class of SUSY
models on observables related to the unitarity triangle (UT) can be
parameterized by two real parameters and by one phase (see
Eqs.~(\ref{f}) and (\ref{g})). In this section we analyze the
implications of this parametrization on the standard analysis of the
UT.  As usual we use the Wolfenstein
parametrization~\cite{wolfenstein} of the CKM matrix in terms of $\l$,
$A$, $\r$ and $\eta$:
\beq V =
\pmatrix{1- {\l^2\over 2} & \l & A \l^3 (\r-i \eta) \cr -\l & 1-
{\l^2\over 2} & A \l^2 \cr A\l^3 (1-\r-i\eta) & -A\l^2 &1\cr } \; .
\label{wolf}
\eeq 
In the following analysis we extend this parametrization beyond
the leading order in $\l$; as a consequence it is necessary to study
the unitarity triangle in the plane $(\bar\rho,\bar\eta)$ where $\bar
\r= \r \;(1-\l^2/2)$ and $\bar \eta=\eta \;(1-\l^2/2)$~\cite{rhobar}.

Let us collect the relevant formulae for $\epsilon_K$, $\D M_{B_{(d,s)}}$ and 
$a_{\psi K_S}$ as functions of $f$, $g$ and $\d_{\tilde c_L \tilde t_2}$:
\beq
  \epsilon_K = -{G_F^2 f_K^2 \hat B_K M_K M_W^2 \over 12 \sqrt{2} \pi^2 \D M_K}  
               \Im \left\{\l_c^{*2} \eta_{cc} C_1^{Wcc} + 2 \l_c^* \l_t^* \eta_{tc} 
               C_1^{Wtc} + \l_t^{*2} \eta_{tt} C_1^{Wtt} (1+f+g) \right\} e^{i\pi\over 4} \; ,
\eeq
\beq
\hskip -6cm
  \D M_{B_d} = -{G_F^2 \over 6 \pi^2} \eta_B M_{B_d}f^2_{B_d}  \hat B_{B_d} M_W^2
             |V_{tb}^{} V_{td}^*|^2 C_1^{Wtt} |1+f+
             g | \; ,  
\eeq
\beq
\hskip -6.5cm \D M_{B_s} = -{G_F^2 \over 6 \pi^2} \eta_B M_{B_s}
f^2_{B_s} \hat B_{B_s}  M_W^2 |V_{tb}^{} V_{ts}^*|^2 C_1^{Wtt} (1+f) \; , 
\eeq
\beq
\hskip -12cm a_{\psi K_S} = \sin 2 (\b + \theta_d) 
\eeq 
where $\l_q = V_{qd}^{} V_{qs}^*$, $\theta_d = {1\over 2} \arg (
1+f+g )$, $\b$ denotes the
phase of $V_{td}^*$ and from \eq{wolf} it follows 
\beq
\sin 2 \b = {2 \bar{\eta} (1-\bar{\rho}) \over (1-\bar{\rho}^2) +
\bar{\eta}^2}\; .  
\eeq
The quantities $\eta_{tt}$, $\eta_{cc}$, $\eta_{cc}$, and $\eta_B$ are NLO
QCD corrections. Their values together with those of the other 
parameters are collected in Table~\ref{hadr}.
\begin{table}
\begin{center}
\begin{tabular}{|c|c|}
\hline 
Parameter & Value \cr \hline\hline
$\eta_{tt}$ & $0.57$ \protect\cite{BJZ1}\cr
$\eta_{cc}$ & $1.38 \pm 0.53$ \protect\cite{Herrlich2} \cr
$\eta_{tc}$ & $0.47 \pm 0.04$ \protect\cite{Herrlich1}\cr
$\hat B_K$ & $0.94 \pm 0.15$ \protect\cite{Draper,Sharpe}\cr 
$\eta_B$ &  $0.55$ \protect\cite{BJZ1}   \cr
$f_{B_d} \sqrt{\hat B_{B_d}}$ & $230 \pm 40 \; \mev$
\protect\cite{Draper,Sharpe,Bernard1} \cr 
$\xi_s={f_{B_s} \sqrt{\hat B_{B_s}}\over f_{B_d} \sqrt{\hat B_{B_d}}}$
 & $1.16 \pm 0.05$ \protect\cite{Bernard2}\cr
\hline 
\end{tabular}
\end{center}
\caption{\it Parameters used in the UT-fits}
\label{hadr}
\end{table}

Our first step is to investigate the regions of the parameter space
spanned by $f$, $g_R$ and $g_I$ that are allowed by the present
experimental data. The procedure consists in writing the 
$\chi^2$ of the selected observables and in accepting only
values of $f$ and $g$ which satisfy
the condition $\min_{\rho,\eta} (\chi^2) \leq 2$. In the
computation of $\chi^2$ we use the following
input
\bea
\epsilon_K &=& (2.271 \pm 0.017) \; 10^{-3}
\mbox{[~PDG ~\protect\cite{Groom}]}, \\
\D M_{B_d} &=& 0.484 \pm 0.010 \; {\rm
ps}^{-1} \mbox{~[HFWG~\protect\cite{HFWG}]} , \\
\left|V_{ub}\over V_{cb}\right|& =& 0.090 \pm 0.025
\mbox{~[PDG  ~\protect\cite{Groom}]}, \\
a_{\psi K_s} &=& 0.79 \pm 0.12 \; . 
\eea
The value quoted above for $\D M_{B_d}$ is the present world average
of experiments on $B$ - $\bar{B}$ mixings, including the recent
BABAR~\cite{Bozzi2001} and BELLE~\cite{Abe2000} measurements.  We
introduce the experimental data on $\D M_{B_s}$ in calculating the
chi--squared ($\chi^2$) using the so--called amplitude
method~\cite{moser}. The prescription consists in adding to the
$\chi^2$ the term $\chi^2_{\D
M_{B_s}}=(1-\mathcal{A})^2/\s_\mathcal{A}^2$ where $\mathcal{A}$ is
the amplitude of the $(B_s-\bar B_s)$ oscillation, given by ($1\pm
\mathcal{A} \cos \D M_{B_s} t$), and $\s_\mathcal{A}$ is the
corresponding error. Both $\mathcal{A}$ and $\s_\mathcal{A}$ are
functions of $\D M_{B_s}$. Notice that using this method, the
statistical interpretation of the value of the $\chi^2$ in its minimum
is preserved. In Ref.~\cite{checchia}, the authors include the $\D
M_{B_s}$ data using an alternative procedure. They consider a
log--likelihood function referenced to $\D M_{B_s} = \infty$ and add
the term $\D \log L^\infty =(1/2 - \mathcal{A})/\s_{ \mathcal{A}}$ to
the $\chi^2$. In this way the significance of the data is
increased. Notice that, in order to interpret the output of this
method in terms of confidence levels it is necessary to perform a
monte--carlo based analysis~\cite{boix}. Since we are interested in
the statistical meaning of the minimum of the $\chi^2$, we prefer to
use the standard amplitude analysis.
 
We present the output of this analysis in \fig{ampl}a. For each
contour we fix the value of $f$ and we require the $\chi^2_{\rm min}$
to be less than 2. Since, as can be seen in \fig{fg}, $f$ is always
smaller than 0.4, we restrict the analysis to $f$ = 0, 0.2 and
0.4. Moreover, for each of these values, we require $|g|$ not to
exceed the upper limit which, according to the analysis of the SUSY
contributions presented in \fig{fg}, we set respectively to 1, 2 and
1.5.  The equation $\sin 2(\b+\theta_d) = a_{\psi K_S}$ has two
solutions (mod $\p$) in which $(\beta+\theta_d)$ lies in the ranges
$[0,\pi/4]$ and $[\pi/4,\pi/2]$ respectively. This implies that, given
a value of $f$, we expect two distinct allowed regions in the
$(g_R,g_I)$ plane, that we call respectively small and large angle
solutions, characterized by $\b+\theta_d < \pi/4$ (i.e.  $\theta_d
\simeq O(0.1)$ and $\b+\theta_d > \pi/4$ (i.e. $\theta_d \simeq
O(1)$). These two regions have some overlap since in the limit
$a_{\psi K_S}=1$ (which is allowed at $\cl{90}$) the two
solutions coincide. We find that, once the upper bound on $|g|$ is
imposed, only a small part of the $f=0.2$ and a tiny corner of the
$f=0.4$ large angle solution survive. Improvements in the experimental
determination of the $B\to X_s \g$ and $B\to \rho \g$ branching ratios
as well as more stringent lower bounds on the SUSY spectrum will have
a strong impact on the allowed $|g|$ values. On the other hand, more
precise measurements of $|V_{ub}/V_{cb}|$, $a_{\psi K_S}$, $\D
M_{B_s}$ and progress in the determination of the relevant hadronic
parameters, will contribute to reduce sizably the size of the allowed
$(g_R,g_I)$ regions. In view of these considerations, we expect the
large angle ($\theta_d \simeq O(1)$) solution is less likely to
survive in future and we will concentrate in the following on the
small angle scenario, i.e., $\theta_d \simeq O(0.1)$.

It is interesting to note that the amplitude method, that is
conservatively used to set the constraint $\D M_{B_s} > 14.9\;
\mbox{ps}^{-1}$ at $\cl{90}$, also yields a 2.5 $\s$ signal for
oscillations around $\D M_{B_s} = 17.7 \; \mbox{ps}^{-1}$.  This
would--be measurement is equivalent to a determination of the ratio
$\D M_{B_s} / \D M_{B_d}$ which in turn depends on the precisely
computed hadronic parameter $\xi_s$: its impact on the unitarity
triangle is thus expected to be quite significant \cite{mfv}. In
\fig{ampl}b we assume this signal to be a measurement with $\D M_{B_s}
= 17.7 \pm 1.4 \; \mbox{ps}^{-1}$ in order to explore its implications
on the previous analysis. The effect of the assumed value of $\D
M_{B_s}$ is to reduce the $g_R,g_I>0$ allowed regions; moreover, its
impact is stronger the higher is the value of $f$. Note, in
particular, that, with the assumed value of $\D M_{B_s}$, the $f=0.4$
contour almost disappeared. This happens because the experimentally
favoured high values of $a_{\psi K_S}$ tend to sharpen the mismatch
between the $\D M_{B_d}$ and $\D M_{B_s}/ \D M_{B_d}$ constraints
(which is due to a non--zero value of $f$).

Before concluding this section we would like to show the impact of the
Extended-MFV model considered in this paper on the profile of the
unitarity triangle in the $(\bar\rho,\bar\eta)$ plane, and the
corresponding profiles in the SM and MFV models.  In Fig.~\ref{cont},
the solid contour corresponds to the SM $\cl{95}$, the dashed one to a
typical MFV case ($f=0.4$, $g= 0$) and the dotted--dashed one to an
allowed point in the Extended-MFV model ($f=0$, $g_R=-0.2$ and
$g_I=0.2$). The representative point that we consider survives all the
experimental constraints examined in the previous section. Using the
values of $\bar\rho$ and $\bar \eta$ that correspond to the central
value of the fit, we obtain the following results for the various
observables: $|V_{ub}/ V_{cb}| = 0.079$, $\epsilon_K = 2.27 \times
10^{-3}$, $\D M_{B_d} = 0.484 \; \mbox{ps}^{ -1} $, $\D M_{B_s} = 20.7
\; \mbox{ps}^{ -1}$ and $a_{\psi K_S} = 0.81$. If $|g|$ is
sufficiently large, $\theta_d$ can be regarded as an essentially free
parameter and the fit will choose the value that gives the best
agreement with the experimental measurement.  In \fig{apsiks} we plot
the $CP$ asymmetry $a_{\psi K_S}$ as a function of $\arg
\delta_{\tilde u_L \tilde t_2}$ (expressed in degrees). The light and
dark shaded bands correspond, respectively, to the SM and the
experimental $1\;\s$ allowed regions. The solid line is drawn for
$f=0$ and $|g| =0.28$.  The experimental band favours $\arg
\delta_{\tilde u_L \tilde t_2}$ in the range
$[0^\circ,100^\circ]$. Employing the explicit dependence
\beq 
\theta_d ={1\over 2}\arg (1+f+ |g| e^{2 i \arg \delta_{\tilde u_L
\tilde t_2}}) \;\;\; (\hbox{mod} \; \pi)\; ,
\eeq
the above phase interval is translated into $ -3^\circ < \theta_d <
8^\circ$, for the assumed values of $|g|$ and f, which is a typical
range for $\theta_d$ for the small angle solution with the current
values of $a_{\psi K_S}$.
\begin{table}
\begin{center}
\begin{tabular}{|c|cc|c|c|c|cc|}
\hline 
Contour & $g_R$ & $g_I$ & $|V_{ub}/V_{cb}|$ & $\D M_{B_s}$ & $a_{\psi K_S}$ & $\a$ & $\g$ \cr \hline
1 & $0.2$  & $0.2$  & 0.094 & 20 $\hbox{ps}^{-1}$ & 0.78 & $119^\circ$ & $40^\circ$ \cr 
2 & $0.0$  & $-0.2$ & 0.110 & 20 $\hbox{ps}^{-1}$ & 0.71 & $101^\circ$ & $51^\circ$ \cr
3 & $-0.4$ & $0.1$  & 0.081 & 17 $\hbox{ps}^{-1}$ & 0.73 & $64^\circ$  & $98^\circ$ \cr \hline
\end{tabular}
\end{center}
\caption{\it Central values of the CKM ratio $|V_{ub}/V_{cb}|$, the
$B_s - \bar B_s$ mass difference $\D M_{B_s}$, the $CP$ asymmetry
$a_{\psi K_S}$ and the inner angles $\a$ and $\g$ of the unitarity
triangle for the contours in the extended--MFV model plotted in
Fig.~\protect\ref{contsamples}. The $(g_R,g_I)$ values for the
contours are also indicated. }
\label{values}
\end{table}

In order to illustrate the possible different impact of this
parametrization on the unitarity triangle analysis, we focus on the
$f=0$ case and choose three extremal points inside the allowed region
in the plane $(g_R,g_I)$. We concentrate on the small angle scenario.
In \fig{contsamples} we plot the $\cl{95}$ contours in the $(\bar
\rho,\bar\eta)$ plane that correspond to the points we explicitely
show in \fig{ampl}a. We summarize in Table~\ref{values} the central
values of $|V_{ub}/V_{cb}|$, $\D M_{B_s}$, the $CP$ asymmetry $a_{\psi
K_S}$, the inner angles $\a$ and $\g$ of the unitarity triangle
computed for the different contours. Contour 1 is drawn for positive
$g_I$ and $\theta_d$ is consequently positive. This implies that
$a_{\psi K_S}$ is expected to be larger than in the SM.  The $CP$
asymmetry and $|V_{ub} / V_{cb}|$ are very close to their world
averages while the angle $\gamma$ is smaller than $\gamma^{SM}$.  In
contour 2 the phase $\theta_d$ is negative and the CP asymmetry is
thus predicted to be lower than the experimental central value and
still $\gamma<\gamma^{SM}$. Contour 3 is drawn for $g=-0.4+0.1 i$ and
is particularly interesting since it corresponds to a solution in
which $a_{\psi K_S}$ is larger than in the SM and the Wolfenstein
parameter $\bar\rho$ is negative, i.e. $\bar\rho<0$, implying a value
of the inner angle $\gamma$ in the domain $\pi/2 < \gamma < \pi$. This
is in contrast with the SM--based analyses which currently yield
$\gamma<\pi/2$ at 2 standard deviations~\cite{mele-ckm, ps-ckm,
bargiotti-ckm, mfv,ciuchini-ckm, buras-ckm, atwood, hocker-ckm} and
with the other solutions shown in \fig{contsamples}. We note that
analyses~\cite{hou,beneke} of the measured two--body non--leptonic
decays $B\to \pi\pi$ and $B\to K \pi$ have a tendency to yield a value
of $\gamma$ which lies in the range $\gamma > \pi/2$ (restricting to
the solutions with $\bar\eta>0$). While present data, and more
importantly the non--perturbative uncertainties in the underlying
theoretical framework do not allow to draw quantitative conclusions at
present, this may change in future. In case experimental and
theoretical progress in exclusive decays force a value of $\gamma$ in
the domain $\pi/2 < \gamma < \pi$, the extended--MFV model discussed
here would be greatly constrained and assume the role of a viable
candidate to the SM.

\begin{table}
\begin{center}
\begin{tabular}{|c|cc|c|c|c|cc|}
\hline 
Contour & $g_R$ & $g_I$ & $|V_{ub}/V_{cb}|$ & $\D M_{B_s}$ & $a_{\psi K_S}$ & $\a$ & $\g$ \cr \hline
1 & $0.2$  & $0.2$  & 0.092 & 20 $\hbox{ps}^{-1}$ & 0.77 & $120^\circ$ & $40^\circ$ \cr 
2 & $0.0$  & $-0.2$ & 0.102 & 20 $\hbox{ps}^{-1}$ & 0.65 & $101^\circ$ & $53^\circ$ \cr
3 & $-0.4$ & $0.1$  & 0.084 & 17 $\hbox{ps}^{-1}$ & 0.75 & $64^\circ$  & $97^\circ$ \cr \hline
\end{tabular}
\end{center}
\caption{\it The same as in Table~\protect\ref{values}. The error on
$|V_{ub}/V_{cb}|$ is reduced by a factor of 2.}
\label{valuesN}
\end{table}
In \fig{contsamplesN} and Table~\ref{valuesN} we show the
consequences, on the analysis described above, of reducing the error
on the CKM ratio $|V_{ub}/V_{cb}|$ by a factor of 2. The value
$|V_{ub}/V_{cb}|=0.090\pm 0.013$, used by us to illustrate the
improved constraint on the CKM unitarity triangle, is obtained from
the inclusive measurement of $|V_{cb}|$ from CLEO~\cite{cleo} and
LEP~\cite{LEP-HFWG}, yielding $|V_{cb}|=(40.57 \pm 1.21) \times
10^{-3}$, and the average of the currently measured values of
$|V_{ub}|$ by the CLEO~\cite{meyer} and LEP~\cite{LEP-HFWG} groups,
yielding $|V_{ub}|=(3.64 \pm 0.46)\times 10^{-3}$. Note that the
reduced error on $|V_{ub}/V_{cb}|$ does not affect sizably the
existence of the $\bar\rho<0$ solution in the extended--MFV model
shown here.

\section{Implications of the Extended-MFV model for $b \ra d\g$
transitions}
In this section we study the implications of the extended--MFV model
on the observables related to the exclusive decays $B \ra \r \g$,
namely the ratio $R (\r \g/K^* \g)$ defined in \eq{ratio}, the isospin
breaking ratio
\beq
\D = {\D^{+0} + \D^{-0} \over 2}\; ,
\eeq
where
\beq
\D^{\pm 0} = { \branch (B^{\pm} \ra \r^\pm \g) \over 
                 2 \branch (B^0 \ra \r^0 \g)} -1 \; ,
\eeq
and the $CP$ asymmetry 
\beq
A_{CP} = {\branch (B^-\ra\r-\g) -\branch(B^+\ra\r^+\g) \over 
          \branch (B^-\ra\r-\g) +\branch(B^+\ra\r^+\g)} \; .
\eeq
We perform the numerical analysis for following set of SUSY input
parameters (that satisfy all the constraints previously discussed):
$\mu=120\;\gev$, $M_2=350\;\gev$, $\tan\beta=4$, $M_{\tilde
t_2}=280\;\gev$, $\theta_{\tilde t}=-0.29$ and $M_{H^\pm}=290\;\gev$.
This allows us to exploit in detail the dependence of the various
observables on the phase of the mass insertion and to give an
illustrative example of the modifications in the profile of these
quantities.
 
In \fig{rdlo} we plot the ratio $R(\r \g/K^* \g)$ as a function of
$\arg \delta_{\tilde t_2 \tilde u_L}$ in the Extended-MFV model, and
compare the resulting estimates with the SM estimates, shown by the
two dashed lines representing the $1\s$ SM predictions.  The solid
curves are the SUSY results obtained for $\r$ and $\eta$ set to their
central values and for $|\d_{\tilde u_L \tilde t_2}|=\d^{\mbox{\tiny
lim}}$, $0.55$ and $0.25$. Here, $\d^{\mbox{\tiny lim}}$ is the
absolute value of the mass insertion that saturates the experimental
upper bound $R<0.28$; it is required to be smaller than 1 and it
depends on the phase of the mass insertion. For the point that we
consider, it varies between 0.6 (for $\arg \delta_{\tilde t_2 \tilde
u_L}=0,\pi$) and 1. The shaded region shown for the $|\d_{\tilde u_L
\tilde t_2}|=\d^{\mbox{\tiny lim}}$ represents the $1\;\s$ uncertainty
in the CKM-parameters ($\bar{\rho},\bar{\eta})$ resulting from the fit
of the unitarity triangle. In the maximal insertion case, the
experimental upper bound $R(\r \g/K^* \g) < 0.28$ is saturated for
$\arg \delta_{\tilde t_2 \tilde u_L} \in [0,\p/2] \cup [3\p/2,2\p]$.
Note that if we require the absolute value of the insertion to be
maximal, the ratio $R(\r \g/K^* \g)$ is always larger than in the
SM. We point out that, in the extended--MFV model, this ratio does not
show a strong dependence on the $\r$ and $\eta$ values as long as
these CKM parameters remain reasonably close to their allowed
region. On the other hand, the impact of reducing $|\delta_{\tilde t_2
\tilde u_L}|$ is quite significant.

Taking into account the discussion at the end of Sec.~\ref{susy} and
that $C_7^s (m_b)$ is negative for all the points that allow for a
sizable mass insertion contribution, the $\arg \delta_{\tilde t_2
\tilde u_L}$ region in which the experimental bound on $R(\r \g/K^*
\g)$ is saturated turns out to be strongly dependent on the sign of
$\overline C_7^{MI}$. Moreover, a large $\overline C_7^{MI}$ is
usually associated with a large stop mixing angle whose sign
determines, therefore, the overall sign of the mass insertion
contribution. In our case, $\overline C_7^{MI} < 0$ and the region
$\Re \delta_{\tilde t_2 \tilde u_L} > 0$ is consequently favoured.
The qualitative behaviour of this plot can be understood rewriting 
\eq{rdelta} as 
\beq 
R(\r\g/K^*\g) = R^{SM} |1 + A
e^{i \arg \delta}|^2 = R^{SM} [1+A^2 + 2 A \cos (\arg \d) ] 
\eeq 
where $A\equiv |\delta_{\tilde t_2 \tilde u_L}| \eta^{16/23} \overline
C_7^{MI} /C_7^s (m_b)$ is positive for the point we consider.

The explicit expressions for the isospin breaking ratio and the $CP$
asymmetry in the SM are~\cite{alibdg}
\bea
\label{dlosm}
\D_{LO} &=& 2 \epsilon_A \left[F_1+{1\over
2}\epsilon_A (F_1^2+F_2^2)\right] \; , \\
A_{CP}  &=& - {2 F_2 (A_I^u - \epsilon_A A_I^{(1)t})\over C_7^{SM} (m_b)
            (1+\D_{LO})} 
\label{acpsm}
\eea
where $\epsilon_A=-0.3$, $A_I^u = 0.046$, $A_I^{(1)t} = -0.016$ and  
\bea
F_1 &=& \Re {V_{ub}^{} V_{ud}^* \over V_{tb}^{} V_{td}^*} \equiv
      - \left| V_{ub}^{} V_{ud}^* \over V_{tb}^{} V_{td}^* \right| \cos \a \; ,\\
F_2 &=& \Im {V_{ub}^{} V_{ud}^* \over V_{tb}^{} V_{td}^*} \equiv
      - \left| V_{ub}^{} V_{ud}^* \over V_{tb}^{} V_{td}^* \right| \sin \a \; . 
\eea
Note that $\epsilon_A$ is
proportional to $1/C_7^{SM} (m_b)$. Eqs.~(\ref{dlosm}) and
(\ref{acpsm}) can be easily extended to the supersymmetric case by
means of the following prescriptions:
\bea
&&V_{td}^* \ra V_{td}^* \; \exp \left\{i \; \arg C_7^d (m_b)\right\} 
  \;\;\;\; [\mbox{i.e.:} \; \a \ra \a - \arg C_7^d (m_b)] \; , \\
&&C_7^{SM} (m_b) \ra |C_7^d (m_b)| \; . 
\eea

In Figs.~\ref{iso} and \ref{acpbdg} we present the results of the
analysis for the isospin breaking ratio and of the $CP$ asymmetry in
$B^\pm \rightarrow \rho^\pm \gamma$, respectively, for the three
representative cases: $|\d_{\tilde u_L \tilde t_2}|=\d^{\mbox{\tiny
lim}}$, $0.55$ and $0.25$ in the Extended-MFV model and compare them
with their corresponding SM-estimates.  Since, in all likelihood, the
measurement of the ratio $R(\r\g/K^*\g)$ will precede the measurement
of either $\Delta$ or the CP-asymmetry in $B\to \rho \g$ decays, the
experimental value of this ratio and the CP-asymmetry $a_{\psi K_S}$
can be used to put bounds on $\vert \delta \vert$ and $\arg\d_{\tilde
u_L \tilde t_2}$. The measurements of $\Delta$ and the CP-asymmetry in
$B^\pm \rightarrow \rho^\pm \gamma$ will then provide consistency
check of this model. Concerning the cases $|\d_{\tilde u_L \tilde
t_2}|=0.55$ and $0.25$, we must underline that large deviations occur
in the phase range in which an unobservably small $R(\r\g/K^*\g)$ is
predicted.  On the other hand, it is interesting to note that, for the
case of maximal insertion and for a phase compatible with the
measurements of $a_{\psi K_S}$ (see \fig{apsiks}), sizable deviations
from the SM can occur for the $CP$ asymmetry but not for the isospin
breaking ratio.

\section{Summary and Conclusions}
The measured $CP$ asymmetry $a_{\psi K_S}$ in $B$ decays is in good
agreement with the SM prediction. It is therefore very likely that
this $CP$ asymmetry is dominated by SM effects. Yet, the last word on
this consistency will be spoken only after more precise measurements
of $a_{\psi K_S}$ and other $CP$--violating quantities are at hand. It
is possible that a consistent description of $CP$ asymmetries in
$B$--decays may eventually require an additional CP-violating phase.
With that in mind, we have investigated an extension of the so--called
Minimal Flavour Violating version of the MSSM, and its possible
implications on some aspects of $B$ physics. The non--CKM structure in
this Extended-MFV model reflects the two non--diagonal mass insertions
from the squark sector which influence the FCNC transitions $b \to d$
and $b \to s$ (see~\eq{delta}). In the analysis presented here, we
have assumed that the main effect of the mass insertions in the
$B$-meson sector is contained in the $b \to d$ transition. This is
plausible based on the CKM pattern of the $b \to d$ and $b \to s$
transitions in the SM. The former, being suppressed in the SM, is more
vulnerable to beyond-the-SM effects. This assumption is also supported
by the observation that the SM contribution in $b \to s \gamma$ decays
almost saturates the present experimental measurements. We remark that
the assumption of neglecting the mass insertion ($\d_{\tilde c_L
\tilde t_2}$) can be tested in the CP-asymmetry in the $b \to s$
sector, such as ${\cal A}_{\rm CP} (B \to X_s \gamma)$, ${\cal A}_{\rm
CP} (B \to K^* \gamma)$, the $B_s^0$ - $\overline{B_s^0}$ mass
difference $\D M_{B_s}$, and more importantly through the induced
CP-asymmetry in the decay $B_s^0(\overline{B_s^0}) \to J/\psi \phi$,
which could become measurable in LHC experiments \cite{ball-lhc} due
to the complex phase of ($\d_{\tilde c_L \tilde t_2}$).  The
parameters of the model discussed here are thus the common mass of the
heavy squarks and gluino ($M_{\tilde q}$), the mass of the lightest
stop ($M_{\tilde t_2}$), the stop mixing angle ($\theta_{\tilde t}$),
the ratio of the two Higgs vevs ($\tan \b_S$), the two parameters of
the chargino mass matrix ($\m$ and $M_2$), the charged Higgs mass
($M_{H^\pm}$) and the complex insertion ($\d_{\tilde u_L \tilde
t_2}$).

We have shown that, as far as the analysis of the unitarity triangle is
concerned, it is possible to encode all these SUSY effects
in the present model in terms of two real parameters ($f$ and
$g_R$) and an additional phase emerging from the imaginary part of $g$ 
($g_I$). We find that despite the inflation of supersymmetric parameters
from one ($f$ in the MFV models) to three (in the Extended-MFV models),
the underlying parameter space can be effectively constrained and the
model remains predictive.
 We have worked out the allowed region in the
plane $(f,|g|=|g_R+i g_I|)$ by means of a high statistic scatter plot
scanning the underlying supersymmetric parameter space, where the
allowed parametric values  are given in Eq.~(\ref{susyparams}). 
The experimental constraints on the parameters emerging from the
branching ratios of the decays $B\ra X_s \g$ and $B\ra \rho \g$
(implemented via the ratio $R(\rho \gamma/K^* \gamma)$), as
well as from the recent improved determination of the magnetic
moment of the muon $(g-2)_\mu$ were taken into account, whereby the
last constraint is used only in determining the sign of the $\mu$-term.

We have done a comparative study of the SM, the MFV-models and the
Extended-MFV model by performing a $\chi^2$-analysis of the unitarity
triangle in which we have included the current world average of the
$CP$ asymmetry $a_{\psi K_S}$ (Eq.~(\ref{sin2betawa})) and the current
lower bound $\D M_{B_s} > 14.9 ~{\rm ps}^{-1}$ using the amplitude
method. Requiring the minimum of the $\chi^2$ to be less than two, we
were able to define allowed-regions in the $(g_R,g_I)$ plane
(correlated with the value of $f$), which are significantly more
restrictive than the otherwise allowed ranges for $|g|$. We studied
the dependence of the CP-asymmetry $a_{\psi K_S}$ on the phase of the
mass insertion ($\d_{\tilde u_L \tilde t_2}$) and find that, depending
on this phase, it is possible to get both SM/MFV-like solutions for
$a_{\psi K_S}$, as well as higher values for the CP-asymmetry.  We
constrain this phase to lie in the range $0^\circ \leq \arg \d_{\tilde
u_L \tilde t_2}\leq 100^\circ $, which typically yields the
Extended--MFV phase $\theta_d$ to lie in the range $-3^\circ \leq
\theta_d \leq 8^\circ$. The assumed measurement of the mass difference
$\D M_{B_s}$, when inserted in the $\chi^2$ analysis, further
restricts the allowed regions in $|g|$. However, as $\D M_{B_s}$ has
not yet been measured, this part of the analysis is mostly
illustrative.  Finally, we have shown the profile of the resulting
CKM--Unitarity triangle for some representative values in the
extended--MFV model. They admit solutions for which $a_{\psi K_S}>
a_{\psi K_S}^{\rm SM}$ and $\gamma> \gamma^{\rm SM}$, favoured by
present data.

To test our model, we have focused on three observables sensitive to
the mass insertion ($\d_{\tilde u_L \tilde t_2}$) related to the
radiative decays $B\ra\r\g$.  We have worked out the consequences of
the present model for the quantities $R(\rho \gamma)/R(K^* \gamma) =
2{\cal B}(B^0 \to \rho^0 \gamma)/{\cal B}(B^0 \to K^{* 0} \gamma)$,
the isospin violating ratio $\Delta^{\pm 0}={\cal B}(B^\pm \to
\rho^\pm \gamma)/2{\cal B}(B^0 \to \rho^0 \gamma) -1$, and direct
CP-asymmetry in the decay rates for $B \to \rho \gamma$ and its charge
conjugate. We conclude that the partial branching ratios in $ B \to
\rho \gamma$, and hence also the ratio $R(\rho \gamma/K^* \gamma)$ can
be substantially enhanced in this model compared to their SM-based
values. The CP-asymmetry can likewise be enhanced compared to the SM
value, and more importantly, it has an opposite sign for most part of
the parameter space. On the other hand, it is quite difficult to
obtain a significant isospin breaking ratio without suppressing the
branching ratios themselves.

Finally, we remark that our analysis has led us to an interesting
observation: the requirement of a positive magnetic moment Wilson
coefficient (i.e., $C_7^s > 0$), entering in $b \to s \gamma$ and $b
\to s \ell^+ \ell^-$ decays, is found to be incompatible with a
sizable contribution to the parameter $g$, which encodes, in the
present model, the non--CKM flavour changing contribution. Thus, it is
possible to distinguish two different scenarios depending on the sign
of $C_7^s$. In the $C_7^s < 0$ case, as in the SM, only small
deviations in the $b \to s$ phenomenology are expected, but sizable
contributions to $g_R$ and $g_I$ are admissible, thereby leading to
striking effects in the $b \to d$ sector. On the other hand, a
positive $C_7^s$ will have strong effects in the $b \to s$ sector but,
since $\vert g \vert$ will be highly constrained, the model being
studied becomes a limiting case of the MFV models.  In particular, in
this scenario, no appreciable change in the CP-asymmetry $a_{\psi
K_s}$ compared to the SM/MFV cases is anticipated. Since the
experimental value of $a_{\psi K_S}$ [\eq{sin2betawa}] is in agreement
with the SM/MFV--models, configurations in which $a_{\psi K_S}$
receives small corrections have a slight preference over the others.
These aspects will be decisively tested in $B$-factory experiments.

\section*{Acknowledgments}
E.L. acknowledges financial support from the Alexander Von Humboldt
Foundation. We would like to thank Riccardo Barbieri, Fred
Jegerlehner, David London, Antonio Masiero and Ed Thorndike for
helpful discussions and communications.

\appendix
\section{Stop and chargino mass matrices}
The $2\times 2$ stop mass matrix is given by
\beq
M^2_{\tilde t} = \pmatrix{ M^2_{\tilde t_{LL}} & M^{2}_{\tilde{t}_{LR}} \cr
                           M^{2}_{\tilde{t}_{LR}} & M^{2}_{\tilde{t}_{RR}} \cr}
\eeq
where
\bea
 M^{2}_{\tilde{t}_{LL}} &=& M^{2}_{\tilde q}
      + ( {1\over 2} - {2\over 3} \sin^{2}\theta_{W} ) \cos 2\beta_S \, M_Z^2 + M_{t}^{2} \enspace , \\
 M^{2}_{\tilde{t}_{RR}} &=& M^{2}_{\tilde q}
      + {2\over 3} \sin^{2}\theta_{W} \cos 2\beta_S \, M_{Z}^{2} + M_{t}^{2} \enspace , \\
 M^{2}_{\tilde{t}_{LR}} &=& M_{t} | A_{t} - \mu^{*} \cot(\beta_S) |
\enspace .
\eea
The eigenvalues are given by
\beq
2 M^{2}_{\tilde t_1, \tilde t_2}
= ( M^{2}_{\tilde{t}_{LL}} + M^{2}_{\tilde{t}_{RR}} )
\pm \sqrt{ ( M^{2}_{\tilde{t}_{LL}} - M^{2}_{\tilde{t}_{RR}} )^{2}
         + 4 ( M^{2}_{\tilde{t}_{LR}} )^{2}}
\enspace ,
\eeq
with $M^2_{\tilde t_2} \le M^{2}_{\tilde t_1}$.
We parametrize the mixing matrix ${\mathcal R}^{\tilde{t}}$ so that
\beq
\pmatrix{ \tilde{t}_{1} \cr \tilde{t}_{2}} =
{\mathcal R}^{\tilde{t}}
\left(\begin{array}{c}
  \tilde{t}_{L} \\ \tilde{t}_{R}
\end{array}\right)
=
\left(\begin{array}{cc} 
 \cos \theta_{\tilde{t}} & \sin \theta_{\tilde{t}} \\
  - \sin \theta_{\tilde{t}} &  \cos \theta_{\tilde{t}}
\end{array}\right)
\left(\begin{array}{c}
  \tilde{t}_{L} \\ \tilde{t}_{R}
\end{array}\right)
\enspace .
\eeq
The chargino mass matrix
\begin{equation}\label{charmass}
M^{\tilde{\chi}^{+}}_{\alpha\beta} =
\left(
\begin{array}{cc}
  M_2                        & M_{W} \sqrt{2} \sin\beta_S  \\
  M_{W} \sqrt{2} \cos\beta_S & \mu
\end{array}
\right)
\end{equation}
can be diagonalized by the bi-unitary transformation
\begin{equation}
\tilde U^{*}_{j\alpha} M^{\tilde{\chi}^{+}}_{\alpha\beta} \tilde V^{*}_{k\beta}
= M_{\tilde{\chi}_{j}^{+}} \delta_{jk}
\enspace ,
\end{equation}
where $\tilde U$ and $\tilde V$ are unitary matrices such that
$M_{\tilde{\chi}_{j}^{+}}$ are positive and
$M_{\tilde{\chi}_{1}^{+}} < M_{\tilde{\chi}_{2}^{+}}$.

\section{Loop functions}
The loop functions for box diagrams, entering in $\varepsilon_K$,
$\Delta M_{B_d}$ and $\Delta M_{B_s}$, are,
\bea
G(a,b) & = & -\frac{ab}{4} \left( {a^2 -8 a + 4\over (a-b)(a-1)^2}\ln{a} 
          \ + \  {b^2 -8 b + 4\over (b-a)(b-1)^2}\ln{b}\ -\ {3\over (a-1)(b-1)} 
          \right) \; ,\nn\\
Y_1(a,b,c,d)& = & {a^2\over (b-a)(c-a)(d-a)}\ln{a} \ +\ {b^2\over (a-b)(d-b)(d-b)}
                  \ln{b}   \nn \\
            &   & +{c^2\over (a-c)(b-c)(d-c)}\ln{c} \ + \ {d^2\over (a-d)(b-d)(c-d)}
                  \ln{d} \; ,\nn\\
Y_2(a,b,c,d) &=&  \sqrt{4 c d}\Bigg[{a\over (b-a)(c-a)(d-a)}\ln {a}\  +\ {b\over 
                  (a-b)(c-b)(d-b)}\ln{b}\nn \\
             & & +{c\over (a-c)(b-c)(d-c)}\ln{c} \  +\ {d\over (a-d)(b-d)(c-d)}
                 \ln{d}\Bigg] \; , \nn \\
Y_1^{MI} (a,b,c,d) &=& {Y_1(a,a,c,d) +Y_1(b,b,c,d) -2 Y_1(a,b,c,d) \over (a-b)^2}
                       \; . \nn
\eea
The loop functions for penguin diagrams, entering in $b \to (s,d)
\gamma$ and in the anomalous magnetic moment of the muon,  are
\begin{eqnarray}
  F_1(x)   &=& {x^3-6x^2+3x+2+6x\ln x\over 12(x-1)^4}\; , \nn \\
  F_3(x)   &=& {x^2-4x+3+2\ln x \over 2(x-1)^3}\; , \nn \\
  f_1(x)  &=& {-7 + 12 x + 3 x^2 - 8 x^3 + 6 x (-2 + 3 x) \log x \over 6 (x-1)^4}\; , \nn\\
  f_2(x) &=& {5 - 12x + 7x^2 - 2x(-2 + 3x)\log x \over 2 (x-1)^3} \; , \nn  \\
  f_{1,2}^{MI}(x,y)  &=& {f_{1,2}(x)-f_{1,2}(y) \over (x-y)}   \; . \nn\\
\end{eqnarray}


\newpage
\begin{figure}[H]
\begin{center}
\epsfig{file=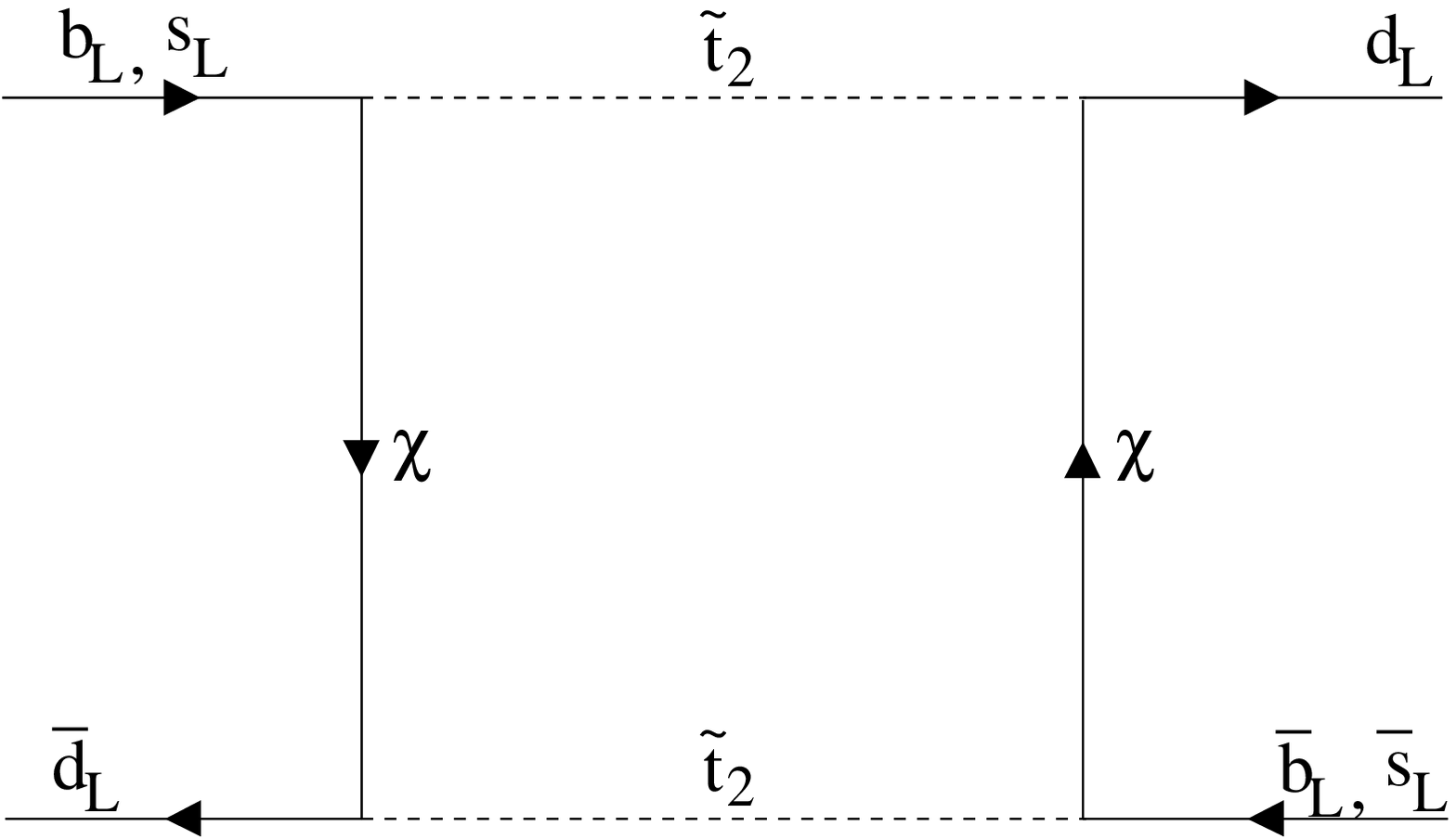,width=0.45\linewidth} 
\quad\quad
\epsfig{file=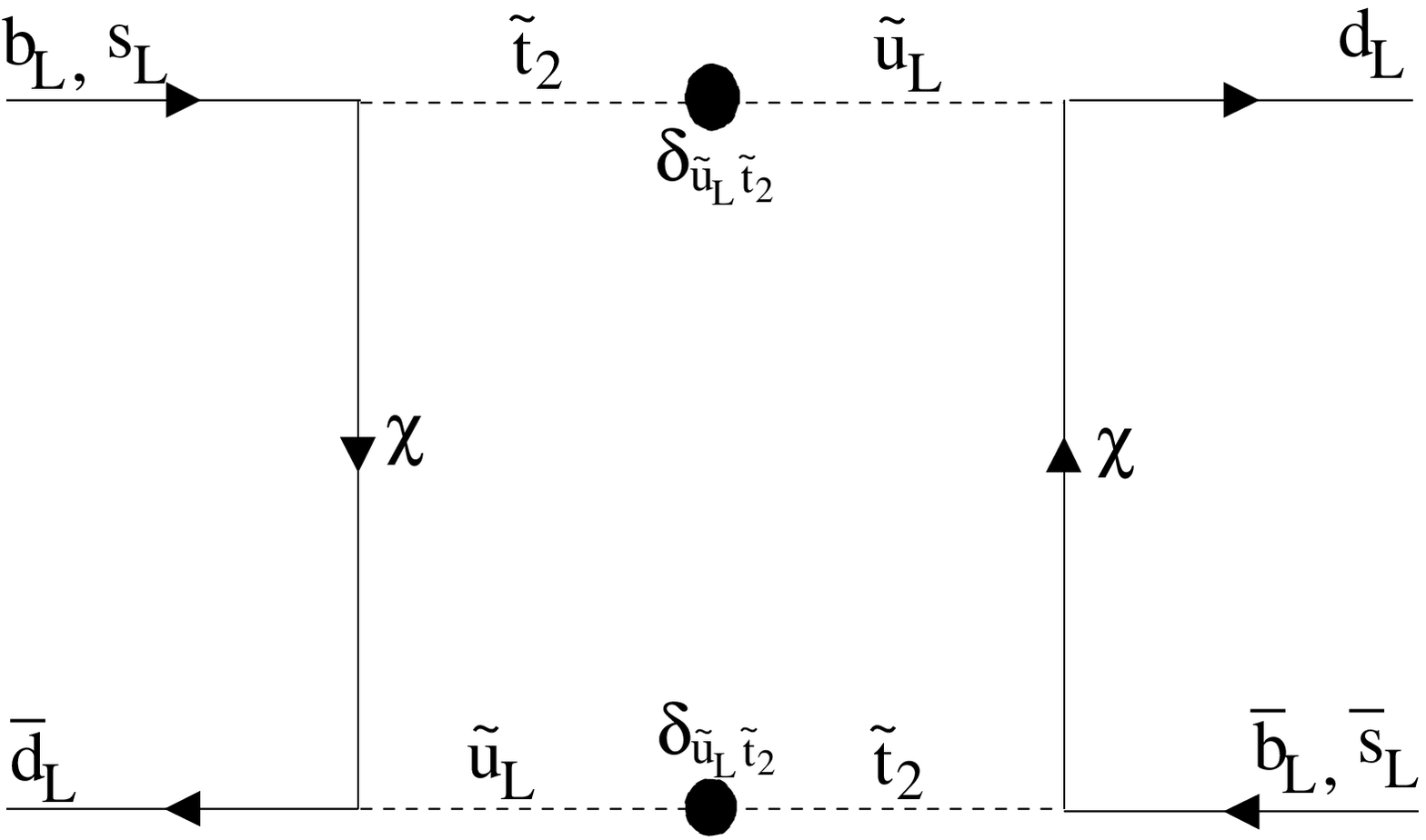,width=0.447\linewidth} 
\caption{\it Feynman diagrams contributing to $C_1^\chi$ and  
$C_1^{MI}$ respectively. The bubble represents the mass insertion
$\d_{\tilde u_L \tilde t_2}$.}
\label{diag}
\end{center}
\end{figure}
\begin{figure}[H]
\begin{center}
\epsfig{file=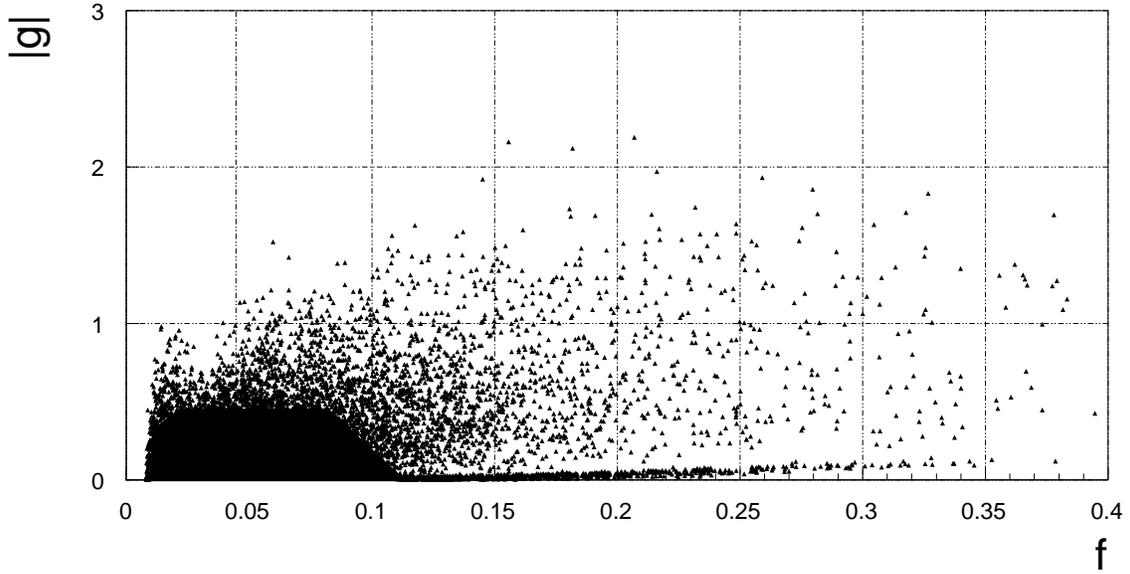,width=0.9\linewidth} 
\caption{\it Allowed points in the $(f,|g|)$ plane. The ranges of the
supersymmetric parameters and the constraints from $b\ra s \gamma$,
$\d a_{\mu^+}$ and $b\ra d \gamma$ that we impose are given in
Eq.~(\protect\ref{susyparams}) and Eq.~(\protect\ref{constraints})
respectively.}
\label{fg}
\end{center}
\end{figure}
\begin{figure}[H]
\begin{center}
\epsfig{file=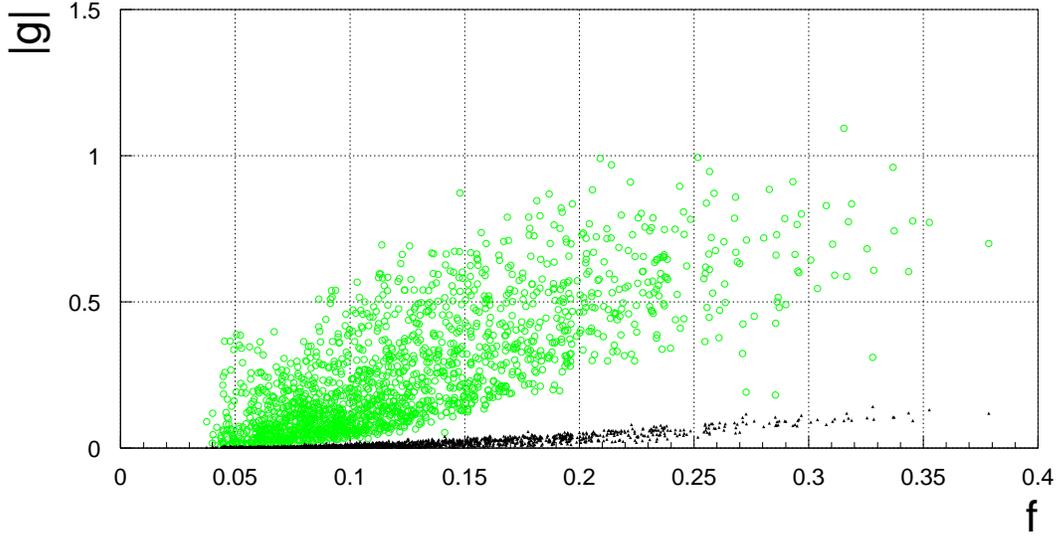,width=0.9\linewidth} 
\caption{\it Allowed points in the $(f,|g|)$ plane that are compatible
with a positive value of the Wilson coefficient $C_7$. The empty circles
satisfy the $b\ra s \gamma$ and $\d a_{\mu^+}$ constraints. The impact of
imposing in addition the upper bound on the $b\ra d \gamma$ transition
is represented by the black dots.}
\label{fgp}
\end{center}
\end{figure}
\begin{figure}[H]
\begin{center}
\epsfig{file=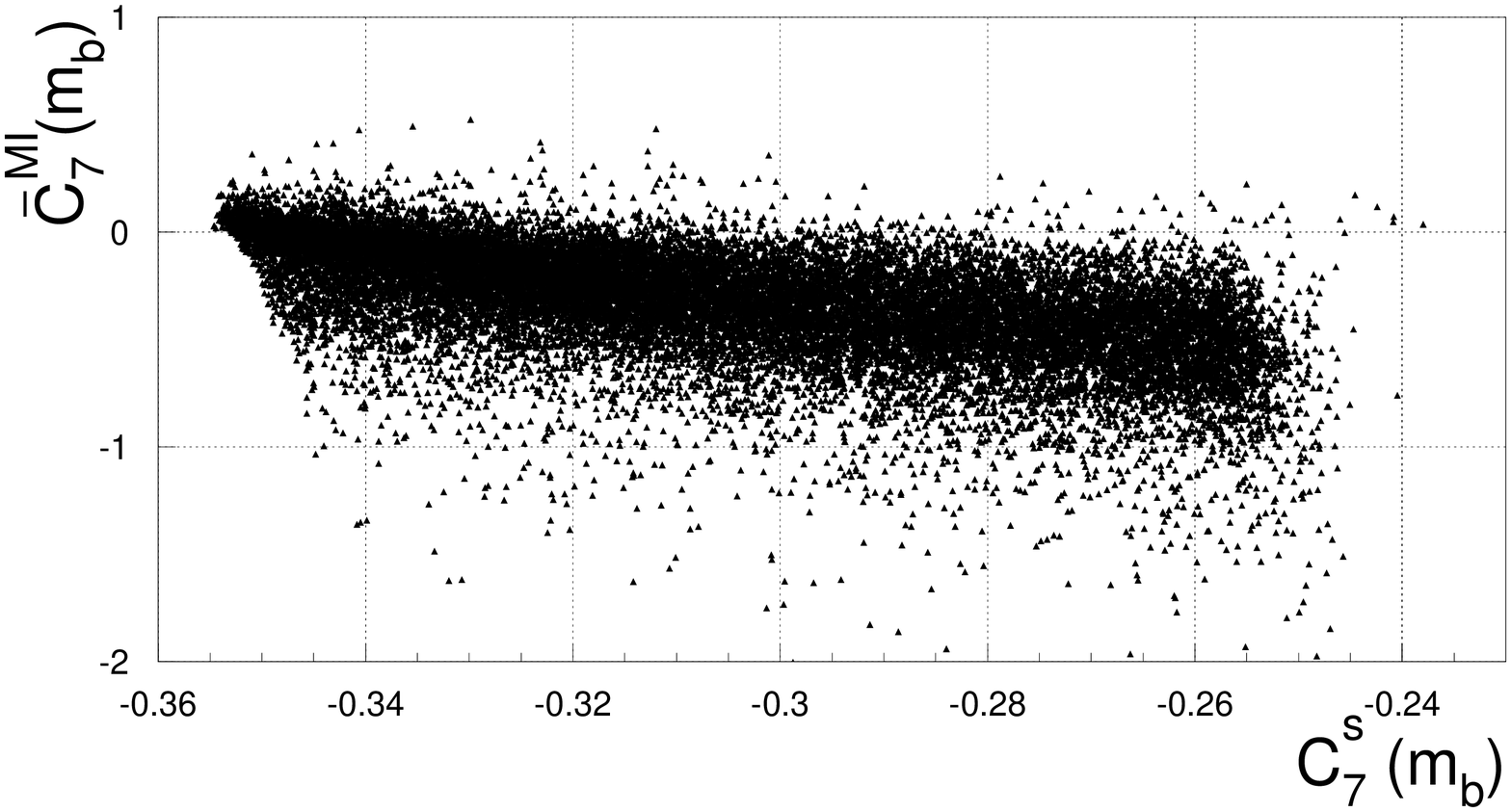,width=0.45\linewidth} 
\epsfig{file=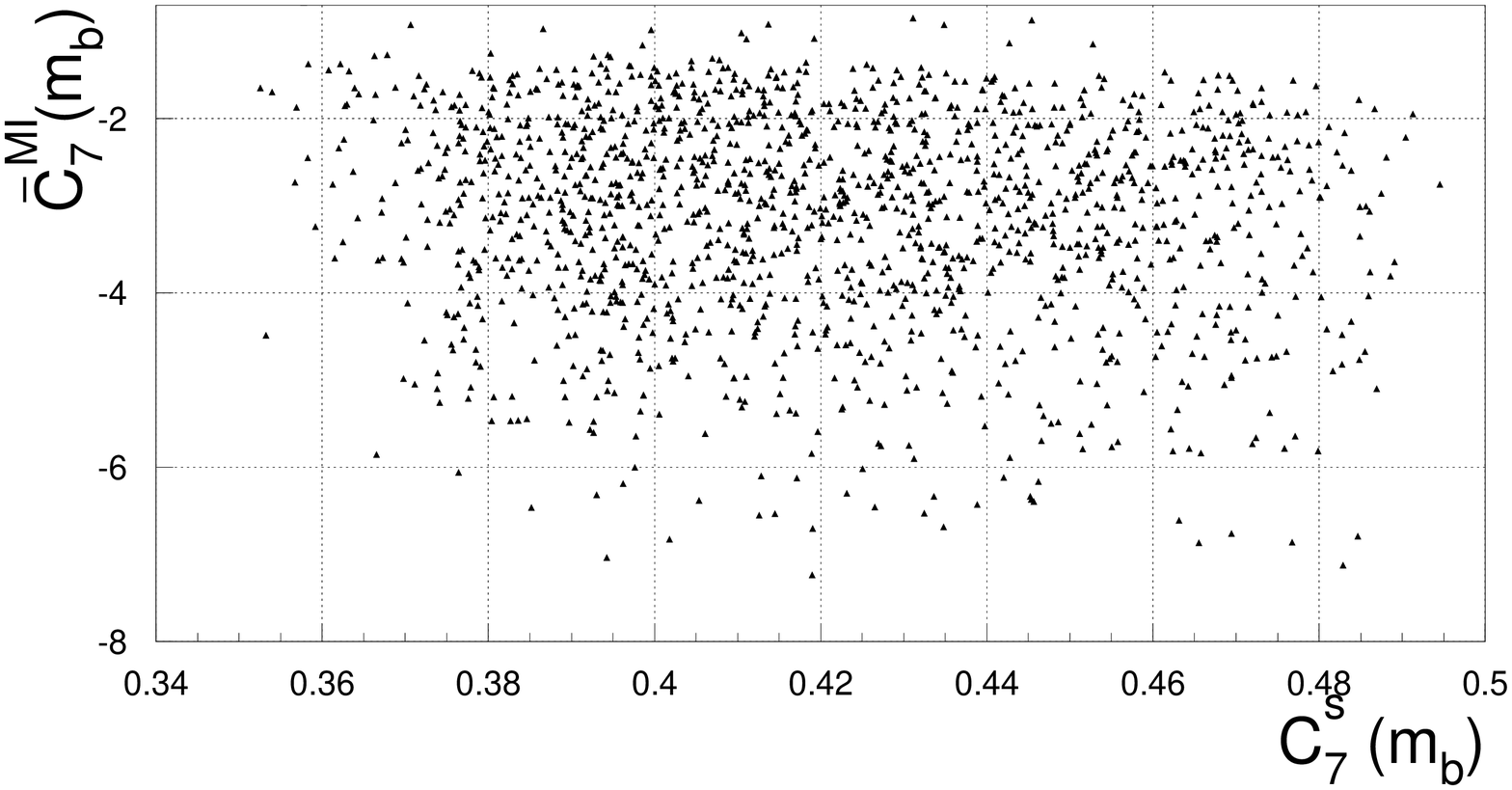,width=0.45\linewidth} 
\caption{\it Correlation between the SUSY contributions to $C_7^s
(m_b)$ and $\overline C_7^{MI}(m_b)$. The plots correspond to the
negative (left figure) and positive (right figure) solutions for
$C_7^s(m_b)$  allowed by the experimental bounds on
$\branch (B\ra X_s \g)$. Note that $C_7^d (m_b)=C_7^s (m_b) +
\overline C_7^{MI} (m_b) \; \d_{\tilde u_L \tilde t_2}$.}
\label{c7c7}
\end{center}
\end{figure}
\begin{figure}[H]
\begin{center}
\epsfig{file=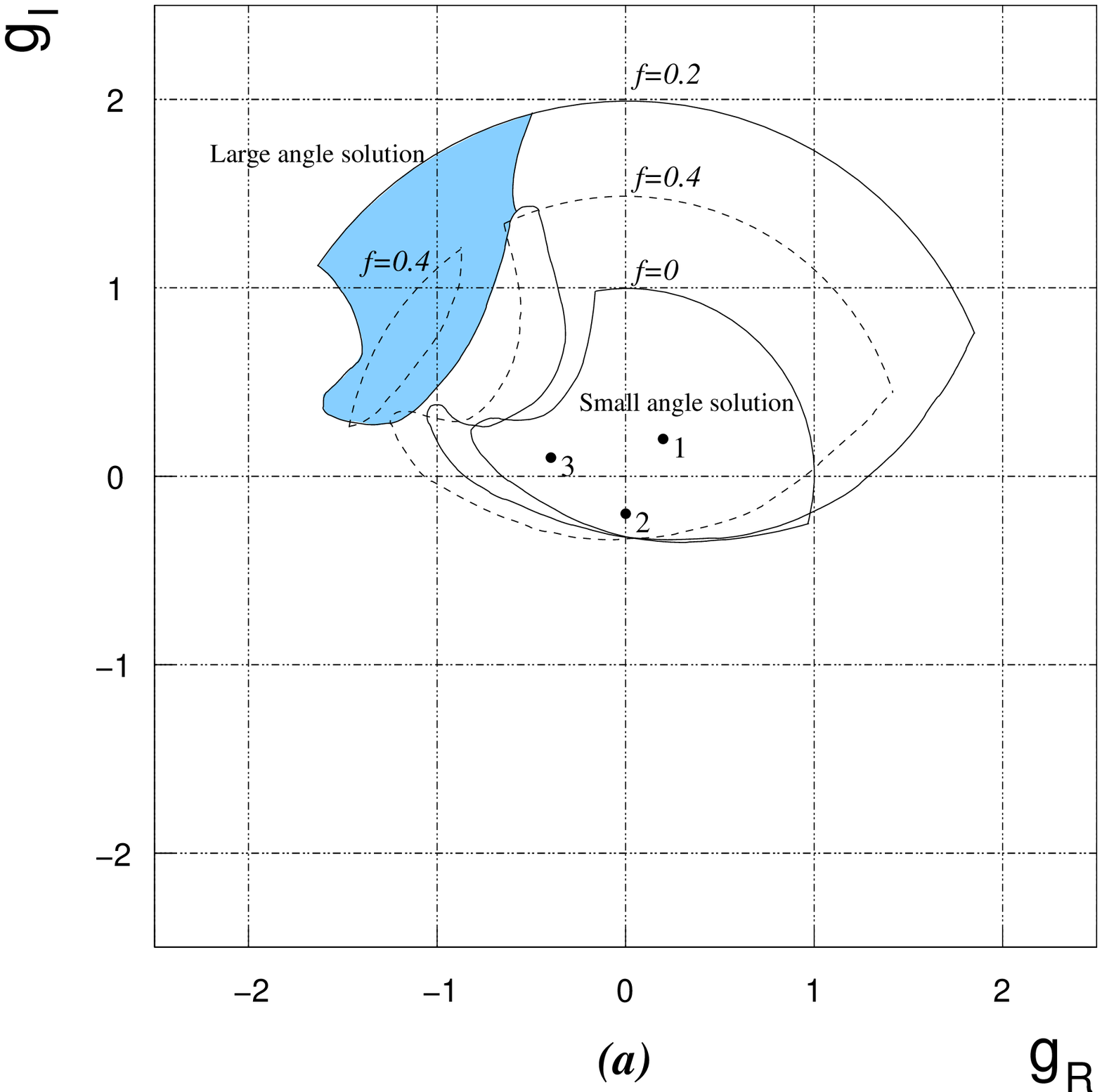,width=0.56\linewidth} 
\epsfig{file=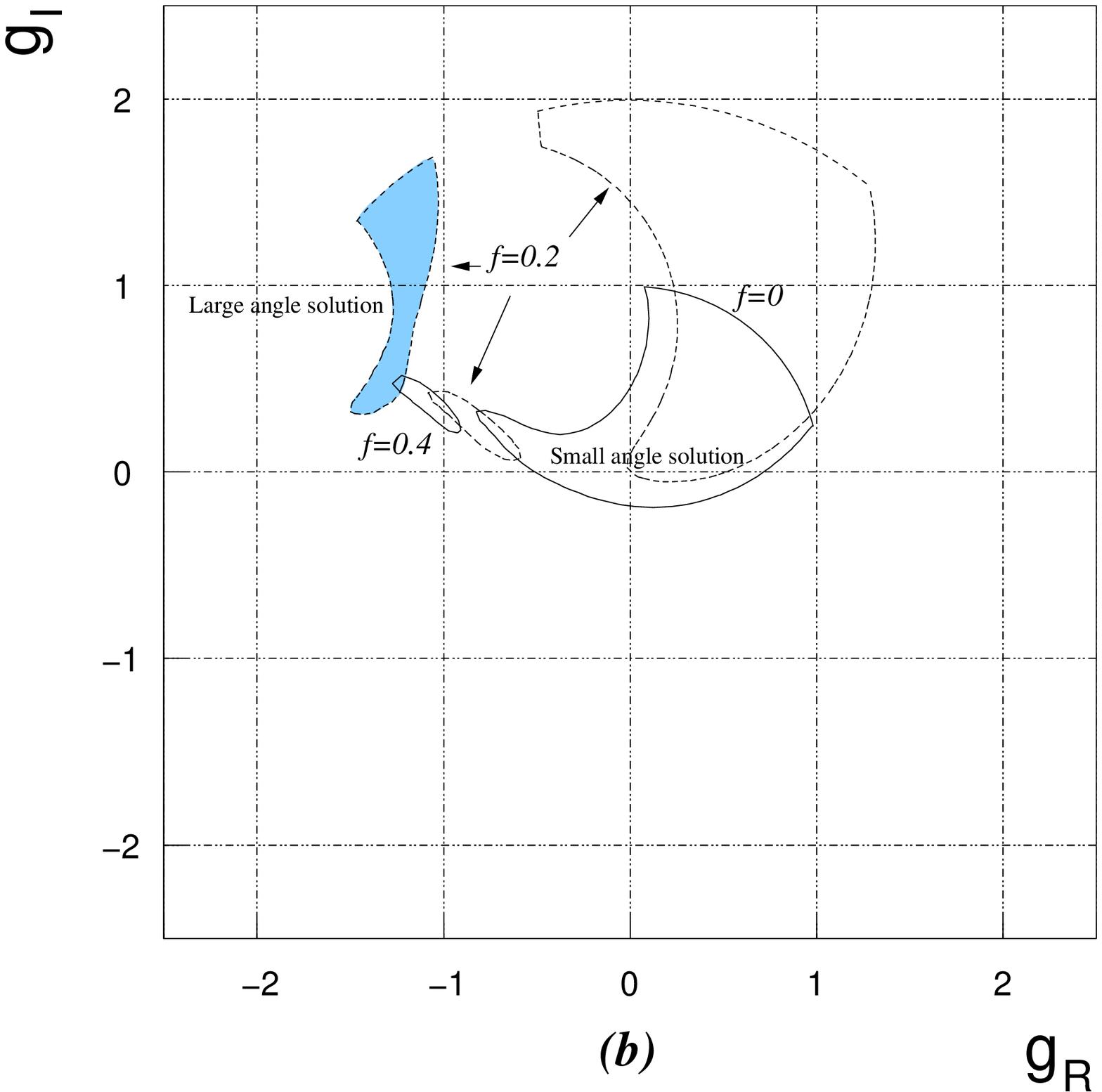,width=0.56\linewidth}
\caption{\it {\bf (a)} Region of the $(g_R , g_I)$ plane for which
$\min_{\rho,\eta} (\chi^2) \leq 2$ from the CKM-UT fits. The $\D
M_{B_s}$ constraint is taken into account using the amplitude method.
The contours correspond to f = 0, 0.2 and 0.4 and the constraints on
$|g|$, coming from Fig.~\protect\ref{fg}, are $|g|\leq 1$, 2 and 1.5
respectively. The shaded areas correspond to a solution with
$\theta+\beta > 45^\circ$ and are allowed for $f=0.2$ and $f=0.4$. The
three dots are representative points that we use to illustrate the
impact of the parametrization we propose on the unitarity triangle.
{\bf (b)} The same as in (a), but interpreting the $2.5 \; \sigma$
enhancement in the amplitude as $\D M_{B_s} = 17.7 \pm 1.4 \;
\mbox{ps}^{-1}$.}
\label{ampl}
\end{center}
\end{figure}
\begin{figure}[H]
\begin{center}
\epsfig{file=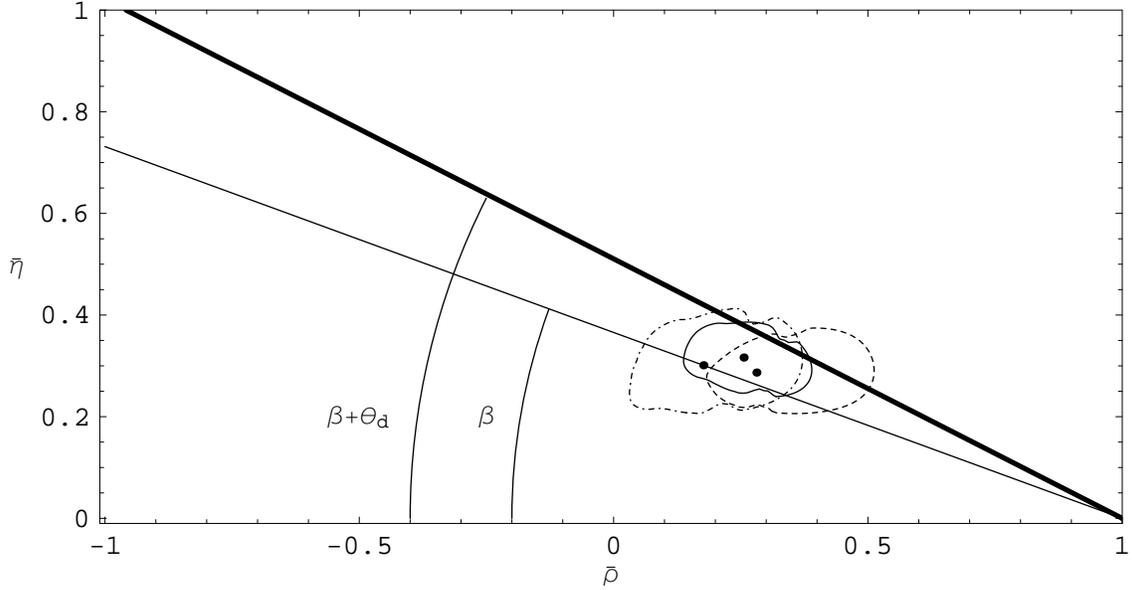,width=0.9\linewidth} 
\caption{\it Allowed $95 \; \% \; C.L.$ contours in the $(\bar\rho,\bar\eta)$
plane.  The solid contour corresponds to the SM case, the dashed contour 
to the Minimal Flavour Violation case with $(f=0.4, \; g=0)$ and the
dashed--dotted contour to the Extended-MFV model discussed in
the text $(f=0, \; g_R=-0.2, \; g_I = 0.2)$.}
\label{cont}
\end{center}
\end{figure}
\begin{figure}[H]
\begin{center}
\epsfig{file=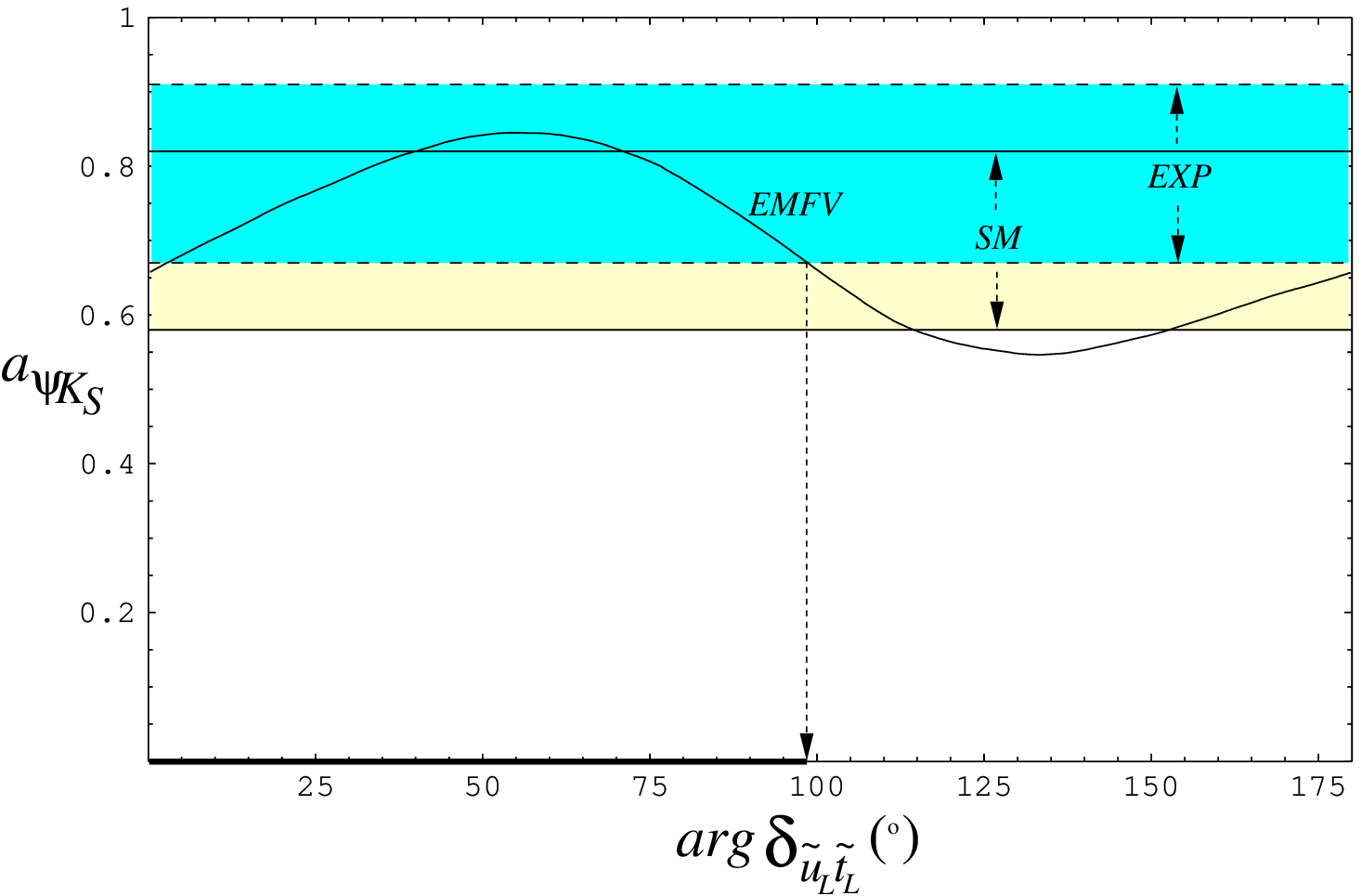,width=0.495\linewidth} 
\epsfig{file=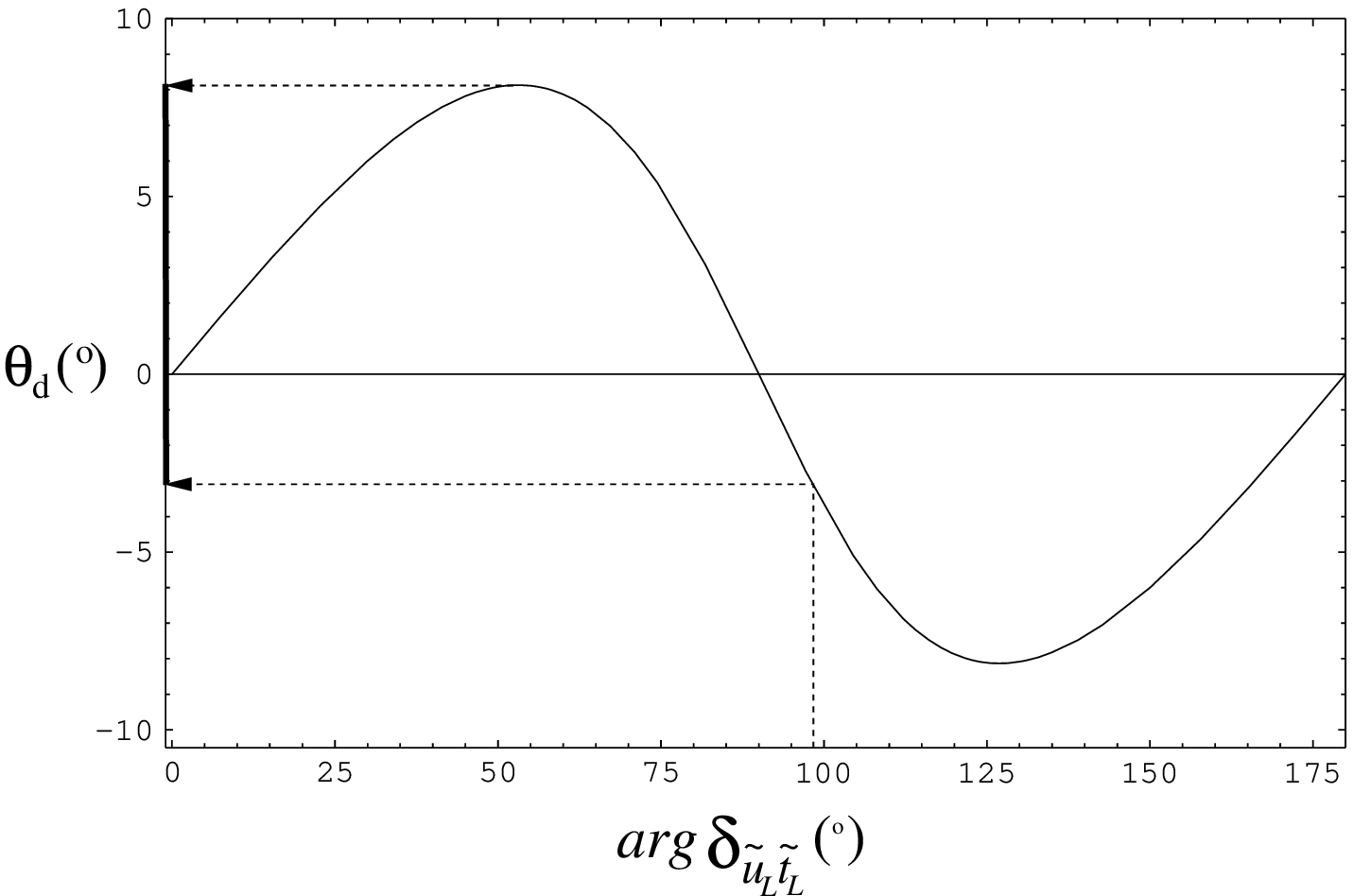,width=0.495\linewidth} 
\caption{\it The $CP$ asymmetry $a_{\psi K_S}$ as a function of $\arg
\delta_{\tilde u_L \tilde t_2}$ expressed in degrees. The solid curve
corresponds to the Extended-MFV model $(f=0, \;|g|=0.28)$. The light
and dark shaded bands correspond, respectively, to the allowed $1\;\s$
region in the SM ($0.58 \leq a_{\psi K_S} \leq 0.82$) and the current
$1 \;\s$ experimental band ($0.67 \leq a_{\psi K_S} \leq 0.91$). The
plot on the right shows the correlation between $\arg \delta_{\tilde
u_L \tilde t_2}$ and the angle $\theta_d$: $\theta_d ={1\over 2}\arg
(1+f+ |g| e^{2 i \arg \delta_{\tilde u_L \tilde t_2}}),\; (\hbox{mod}
\; \pi)$. The experimentally allowed region favours $0^\circ < \arg
\delta_{\tilde u_L \tilde t_2}< 100^\circ$ that translates into $
-3^\circ < \theta_d < 8^\circ$.}
\label{apsiks}
\end{center}
\end{figure}
\begin{figure}[H]
\begin{center}
\epsfig{file=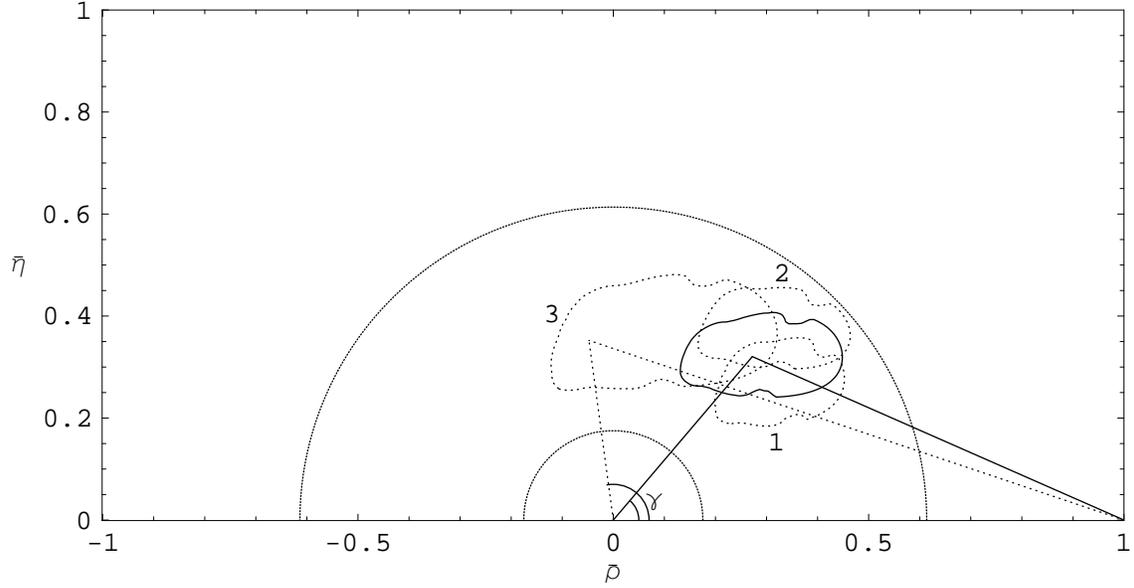,width=0.9\linewidth} 
\caption{\it Allowed $\cl{95}$ contours in the $(\bar\rho,\bar\eta)$
plane. The solid contour is the SM case. The two semicircles represent
the $2\; \sigma$ region allowed by $|V_{ub}/V_{cb}|=0.090\pm 0.025$.
The contours numbered 1 to 3 correspond respectively to the points
indicated in \fig{ampl}a, and their $(g_R,g_I)$ values are given in
Table~\ref{values}. Note that the UT--contour 3 yields
$\gamma>\pi/2$.}
\label{contsamples}
\end{center}
\end{figure}
\begin{figure}[H]
\begin{center}
\epsfig{file=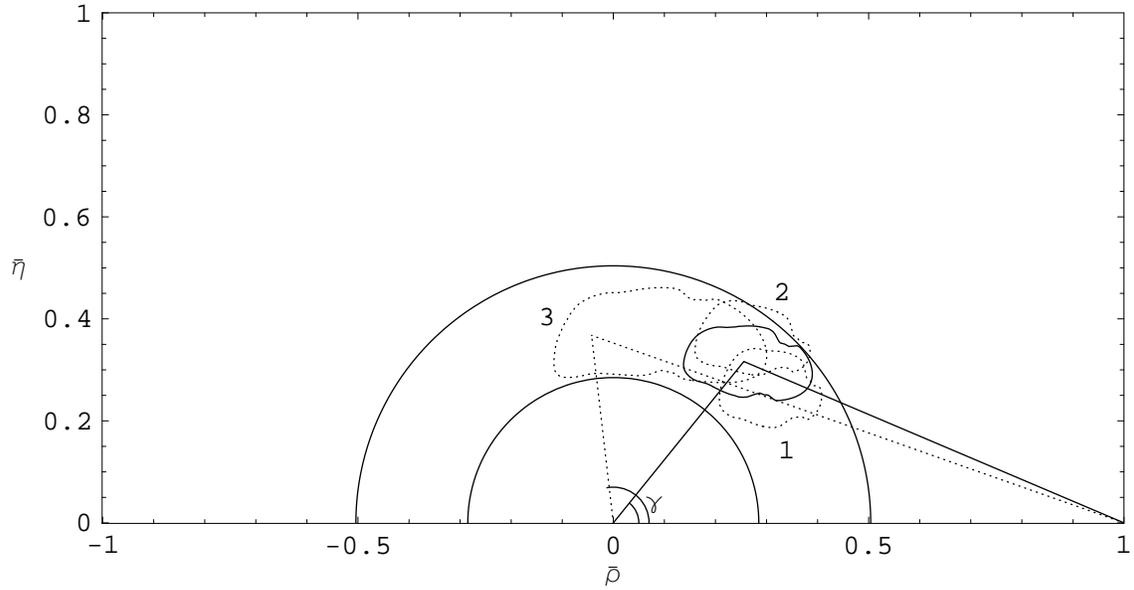,width=0.9\linewidth} 
\caption{\it The same as in Fig.~\protect\ref{contsamples}. The error
on $|V_{ub}/V_{cb}|$ is reduced by a factor of 2. The two semicircles
represent the $2\; \sigma$ region allowed by $|V_{ub}/V_{cb}|=0.090\pm
0.0125$. The $(g_R,g_I)$ values corresponding to the three
extended--MFV contours are given in Table~\ref{valuesN}.}
\label{contsamplesN}
\end{center}
\end{figure}
\begin{figure}[H]
\begin{center}
\epsfig{file=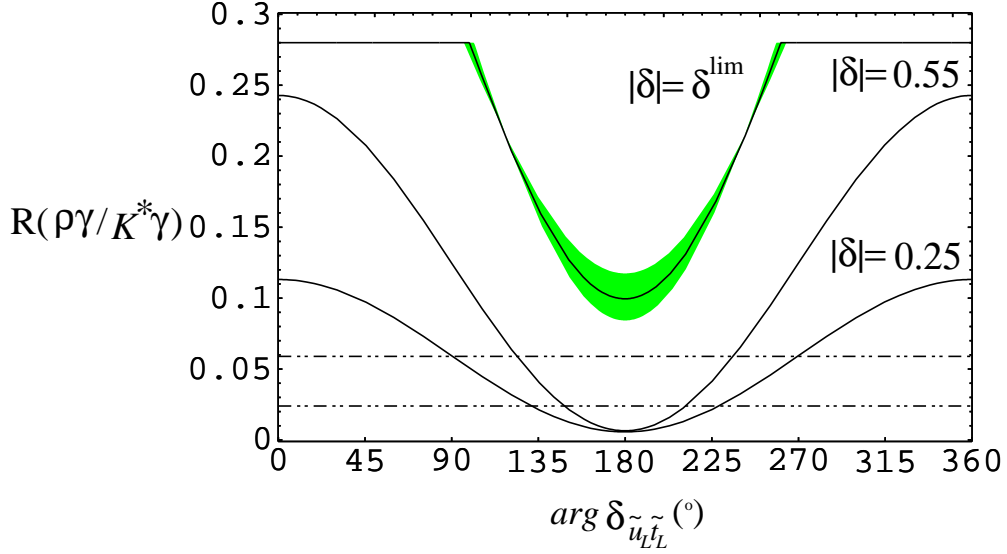,width=0.8\linewidth} 
\caption{\it The ratio $R (\rho \gamma/K^* \gamma)= \branch (B
\rightarrow \rho \gamma) / \branch (B \rightarrow K^* \gamma)$ as a
function of $\arg \delta_{\tilde t_2 \tilde u_L}$ (in degrees) in the
Extended-MFV model, satisfying the present experimental upper bound $R
(\rho \gamma/K^* \gamma) < 0.28$ (at 90\% C.L.). The solid lines are
obtained for $\bar{\r}$ and $\bar{\eta}$ set to their central values
and for $|\d_{\tilde u_L \tilde t_2}|=\d^{\mbox{\tiny lim}}$, $0.55$
and $0.25$. The shaded region in the top curve represents the $1 \;\s$
uncertainty due to the fit of the unitarity triangle. The dashed lines
indicate the $1 \;\s$ SM prediction.}
\label{rdlo}
\end{center}
\end{figure}
\begin{figure}[H]
\begin{center}
\epsfig{file=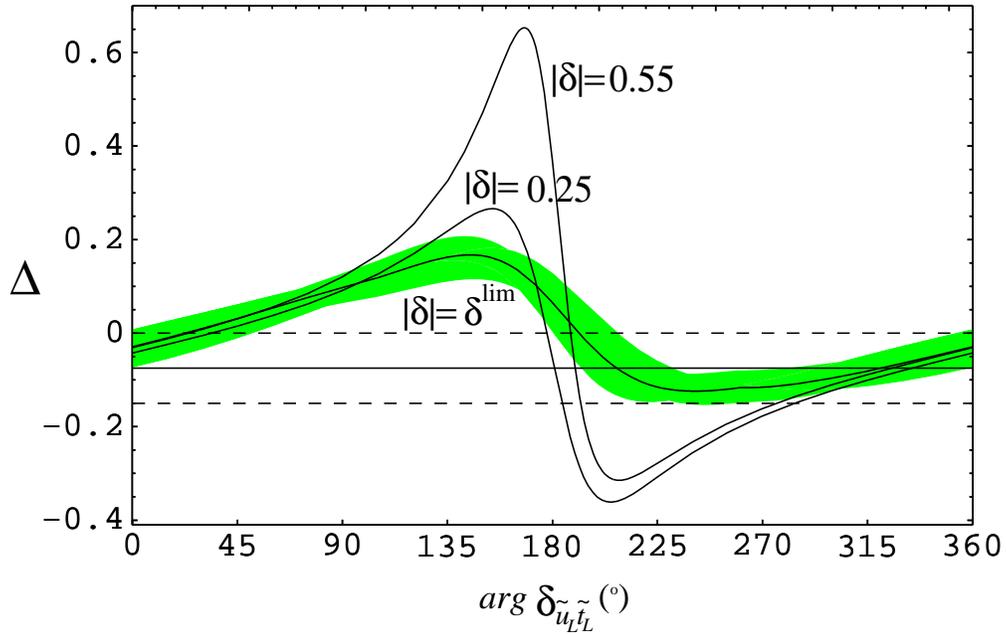,width=0.8\linewidth} 
\caption{\it The isospin breaking ratio $\Delta$ as a function of
$\arg \delta_{\tilde t_2 \tilde u_L}$ (in degrees).  See the
caption in \fig{rdlo} for further explanations.}
\label{iso}
\end{center}
\end{figure}
\begin{figure}[H]
\begin{center}
\epsfig{file=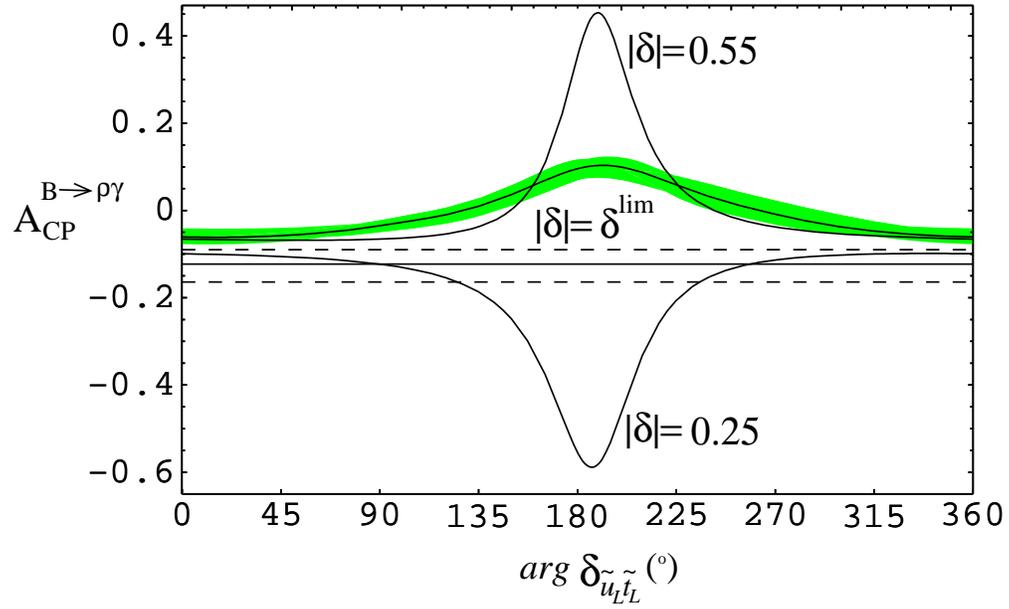,width=0.8\linewidth} 
\caption{\it The $CP$ asymmetry in $B^\pm \rightarrow \rho^\pm \gamma$
as a function of $\arg (\delta_{\tilde t_2 \tilde u_L})$ (in
degrees). See the caption in Fig.~\ref{rdlo} for further
explanations.}
\label{acpbdg}
\end{center}
\end{figure}

\end{document}